\documentclass[aps,pra,twocolumn,amsmath,amssymb,showpacs,superscriptaddress,nobalancelastpage]{revtex4-1}
\pdfoutput=1
\usepackage{graphicx}
\usepackage{dcolumn}
\usepackage{bm}
\usepackage{graphicx}
\usepackage{xcolor}
\usepackage{url}
\usepackage[normalem]{ulem}
\usepackage{times}
\usepackage{bm}
\usepackage{mathrsfs} 
\usepackage{dsfont} 
\usepackage[colorlinks=true,breaklinks=true,allcolors=blue]{hyperref}
\usepackage{braket}

\begin{document}
\title{A versatile quantum walk resonator with bright classical light} 

\author{Bereneice Sephton}
\affiliation{School of Physics, University of the Witwatersrand, Private Bag 3, Wits 2050, South Africa} 
\affiliation{CSIR National Laser Centre, PO Box 395, Pretoria 0001, South Africa} 
\author{Angela Dudley}
\affiliation{School of Physics, University of the Witwatersrand, Private Bag 3, Wits 2050, South Africa} 
\affiliation{CSIR National Laser Centre, PO Box 395, Pretoria 0001, South Africa} 
\author{Gianluca Ruffato}
\affiliation{Department of Physics and Astronomy G. Galilei, University of Padova, via Marzolo 8, 35131 Padova, Italy}
\author{Filippo Romanato}
\affiliation{Department of Physics and Astronomy G. Galilei, University of Padova, via Marzolo 8, 35131 Padova, Italy}
\affiliation{CNR-INFM TASC IOM National Laboratory, S.S. 14 Km 163.5, 34012 Basovizza, Trieste, Italy}
\author{Lorenzo Marrucci}
\affiliation{Dipartimento di Fisica, University di Napoli Federico II, Complesso Universitario di Monte S. Angelo, via Cintia, 80126 Napoli, Italy}
\author{Miles Padgett}
\affiliation{SUPA, School of Physics and Astronomy, University of Glasgow, Glasgow, G12 8QQ, UK}
\author{Sandeep Goyal}
\affiliation{Indian Institute of Science Education and Research, Mohali, Punjab 140306, India}
\author{Filippus Roux}
\affiliation{School of Physics, University of the Witwatersrand, Private Bag 3, Wits 2050, South Africa} 
\affiliation{National Metrology Institute of South Africa, Meiring Naude Road, Pretoria, South Africa}
\author{Thomas Konrad}
\affiliation{School of Chemistry and Physics, University of KwaZulu-Natal, Private Bag X54001, Durban 4000, South Africa}
\author{Andrew Forbes}
\email[Corresponding author: ]{andrew.forbes@wits.ac.za}
\affiliation{School of Physics, University of the Witwatersrand, Private Bag 3, Wits 2050, South Africa}



\begin{abstract}
\noindent In a Quantum Walk (QW) the "walker" follows all possible paths at once through the principle of quantum superposition, differentiating itself from classical random walks where one random path is taken at a time. This facilitates the searching of problem solution spaces faster than with classical random walks, and holds promise for advances in dynamical quantum simulation, biological process modelling and quantum computation. Current efforts to implement QWs have been hindered by the complexity of handling single photons and the inscalability of cascading approaches. Here we employ a versatile and scalable resonator configuration to realise quantum walks with bright classical light. 
We experimentally demonstrate the versatility of our approach by implementing a variety of QWs, all with the same experimental platform, while the use of a resonator allows for an arbitrary number of steps without scaling the number of optics. Our approach paves the way for practical QWs with bright classical 
light and explicitly makes clear that quantum walks with a single walker do not require quantum states of light.
\end{abstract}
\maketitle

\section{Introduction}

\noindent  Quantum walks (QWs) may arguably be dated back to Feynman and even Dirac \cite{wang2013physical}, but certainly have become topical since the seminal paper by Aharonov et.\ al.\ \cite{aharonov1993quantum}, offering a probability distribution that spreads quadratically faster than that of a classical random walk.  In a typical random walk a discrete walker is randomly moved in directions, say left or right, indicated by the outcome of a random value generator (coin toss). Unlike in the case of the classical random walk, a superposition of coin and position states occur for all possibilities in the QW, that is, all possible paths are walked at each step in the process. This changes the dynamics of propagation, resulting in interference of the probability amplitudes for each position held by the walker and inducing a ballistic speedup in the probability distribution variance over the classical equivalent. The interest in QWs stems partly from the realisation that they provide a universal basis for quantum computation \cite{childs2009universal}, presenting an alternate route for successful development of a feasible quantum computer.  Consequently, a large variety of QW algorithms have been designed, allowing database searching \cite{childs2004spatial, shenvi2003quantum}, navigation of networks \cite{sanchez2012quantum}, quantum simulation \cite{venegas2012quantum, cardano2014photonic, schreiber20122d} and element distinctness determination, \cite{ambainis2007quantum} among others. Recently, its application has been extended into quantum cryptography with image encryption \cite{yang2015novel} and quantum key distribution \cite{vlachou2017quantum} protocols. Studies into photosynthetic energy transport in systems also show the natural occurrence of QWs \cite{mohseni2008environment}, indicating its promise as a modeling technique for certain natural processes.

Inevitably there has been much attention on how to actually implement a QW, which have now been demonstrated with many systems including Nuclear Magnetic Resonance (NMR) \cite{ryan2005experimental}, electrons \cite{feist2015quantum}, atoms \cite{karski2009quantum}, ions \cite{zahringer2010realization, schmitz2009quantum}, photons \cite{cardano2014photonic}, Bose-Einstein condensate \cite{alberti2017quantum}, optical fiber time loops \cite{schreiber2011decoherence, boutari2016large}, OAM \cite{cardano2014photonic}, photonic waveguide arrays \cite{perets2008realization, sansoni2012two} and cascaded q-plates \cite{cardano2015quantum, cardano2016statistical}, all in the quantum regime.  Many authors claim to attenuate their laser pulses, ostensibly to reach the single photon regime so that they may claim a ``quantum'' walk.  Such set-ups have included optical cavities \cite{bouwmeester1999optical}, photonic crystal chips \cite{qi2016experimentally}, and optical fiber time loops \cite{regensburger2011photon}. In these approaches the walker is directed through physical paths involving multiple interferometers to achieve the interference effect, without the use of a coin degree of freedom, akin to a Galton board. Here cascaded q-plates \cite{cardano2017detection} have also been employed, avoiding this issue, however still subject to scalability and measurement issues. 

Using a coin makes it possible to implement QWs without interferometers. For this purpose, entanglement (non-separable correlation) between the coin and position degrees-of-freedom (DoF) is a necessary requirement. In this sense, coined QWs may be considered to act as entanglement generators \cite{venegas2008quantum}, and the use of classical light to achieve this has been outlined theoretically \cite{goyal2013implementing}. But non-separability is not unique to quantum mechanics and has been observed in many classical systems, including the non-separable correlations in local DoFs of vector beams \cite{borges2010bell, kagalwala2013bell, ghose2014entanglement, lee2004entanglement}. These correlations are known as ``classical non-separability" or ``classical entanglement" \cite{spreeuw1998classical}.  

\begin{figure}[htbp]
	\centering
	\includegraphics[width=\linewidth]{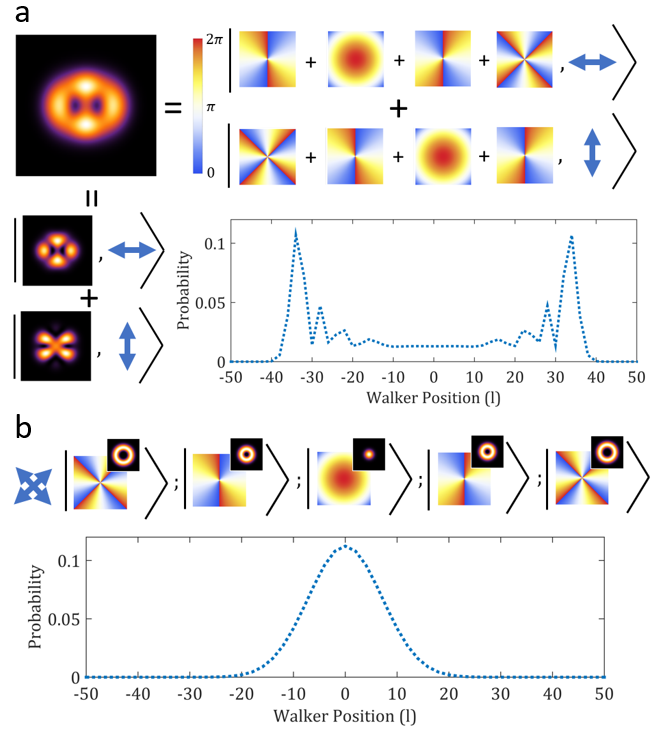}
	\caption{\textbf{Classically entangled light.} Illustration of the QW reliance upon the non-separability of OAM and polarization states of classical light. (a) shows the non-separable superposition of OAM states and polarization for a 4 step QW. False color density plots show the phase distribution of the OAM modes \textit{l} = \{-4,-2, 0, 2, 4\}, forming the positions occupied by the walker (light beam). Weighted superpositions of these would then yield the spatial modes shown in the bottom-left corner with the associated polarization states. Together, the spatial distribution at the top right is formed, where the QW is contained in a single beam of classically entangled light. After many steps, the walker occupies positions near the end of the distribution with greater probability as is shown for 50 steps of a symmetrical walk at the bottom left corner. (b) shows the 4-step classical RW analogue where the walker may only assume one position (OAM) state at a time. Here, the state of the walker remains separable. Accordingly, the light beam (walker) contains only one of the possible OAM states shown as one of the kets with a separable polarization state. Insets show the spatial distribution of the beam for each case. Measurements of the repeated walk then results in the probability distribution below after 50 steps with the walker occupying positions at the center with the greatest probability.}
	\label{fig:twists}
\end{figure}

Here we experimentally demonstrate a versatile and scalable resonator configuration to realise quantum walks with bright classically entangled light. We use vector vortex beams and geometric phase to demonstrate a quantum walk that takes place with spatial modes of classical light in orbital angular momentum (OAM) space where polarization acts as the internal coin state. The versatility of our approach allows us to implement a variety of QWs, for example, symmetric and asymmetric QWs, transitions from a Hadamard coin to a balanced coin QW and a NOT-coin QW, all with the same experimental platform. Our resonator-type configuration, which is implemented by means of a ring cavity, overcomes scalability and flexibility issues associated with many cascaded step schemes, where the resources scale linearly with the number of steps.  This experimental demonstration of a QW  makes explicit the equivalence of non-separable states, classical and quantum, in so far as a QW is concerned.  We do not attenuate the light, instead using bright classical light to implement our QW.  Consequently, we can scarcely avoid the conclusion that quantum walks with a single walker do not require quantum states of light.

\section{Results}
\noindent \textbf{Theory.}  Like in a classical random walk, each step of a quantum walk  consist of a change of the coin state (corresponding to a coin toss in a random walk) and a shift of the position of a walker according to the coin state. Our discrete time quantum walk of $n$ steps on a one-dimensional lattice requires  a system with two degrees of freedom. The first degree of freedom ({\it DoF}) has to be associated with a linear state space that comprises a  number $2n +1$ of orthogonal states and the second DoF needs at least two orthogonal states. In our case we choose a laser pulse with Gaussian profile carrying OAM ($l$) that can assume the values  $l=-n, -n+1, \ldots n$. It is well known that the corresponding optical Laguerre Gauss modes correspond to elements of a Hilbert space of  square-integrable functions and can be represented in Dirac notation by the vectors $\ket{l=-n}, \ket{l=-n-1}, \ldots \ket{l=n}$. The polarisation of the light corresponds to the second degree of freedom spanned by left-circular and right circular basis states, represented by the Dirac 'ket vectors' $\ket{L}$ and $\ket{R}$, respectively. Here we consider the states from the point of view of the source.

A move of the walker to the left  or right, given that the coin shows the value $L$ or $R$, respectively, can then be expressed by a shift operator $\hat{S}$ acting on the OAM and polarisation states of the light pulse:  
\begin{equation}
\hat{S} = \sum^{\infty}_{l = -\infty} \ket{l+2q}\bra{l} \otimes \ket{L}\bra{R} + \ket{l-2q}\bra{l} \otimes \ket{R}\bra{L}.
\label{eqn:shift}
\end{equation}
This addition and subtraction of the value $2q$ depending on the polarisation, corresponding to a translation ({\it 'walk'} ) in OAM space can be realised using a so-called q-plate (cp.\ description of the experiment).
Note that the action of the q-plate described by the shift operator $\hat{S}$ includes an additional flip of the polarisation, which corresponds to a change of the coin state. This effect has to be taken into account when defining the coin toss operator. We  implemented several coin toss operators (see below). All of these can be easily realised by means of quarter-wave plates (QWP), 
\begin{align}
    \hat{Q}_{\theta}= \frac{1}{\sqrt{2}} & \left(\ket{R}\bra{R} + i \exp(-i 2\theta)\ket{R}\bra{L}\right.\nonumber\\ 
    &\left.  + i \exp(i 2\theta)\ket{L}\bra{R}+ \ket{L}\bra{L}\right)\,, 
    \label{genqwp}
\end{align}
and half-wave plates (HWP), 
\begin{equation}
    \hat{H}_{\theta}=  i \exp(-i 2\theta) \ket{R}\bra{L} + i \exp(i 2\theta)\ket{L}\bra{R}\,, 
    \label{genhwp}
\end{equation}
where $\theta$ is the angle of the fast axis of the plates to the horiziontal axis.
For example, a QWP with a fast-axis angle of $45^{\circ}$ combined with a q-plate results in a QW with a Hadamard coin specified by the propagation operator
\begin{eqnarray}
\hat{Z}_H\equiv\hat{Q}_{45^{\circ}} \hat{S} = \sum_l\left( \ket{l+2q}\bra{l}\otimes\frac{1}{\sqrt{2}}(\ket{R}+\ket{L}) \bra{R}\right.\nonumber\\ + \left. \ket{l-2q}\bra{l}\otimes\frac{1}{\sqrt{2}}(\ket{R}-\ket{L}) \bra{L} \right)\,. 
\label{eqn:QW}
\end{eqnarray}

\begin{figure*}
\centering
	\includegraphics[width=\linewidth]{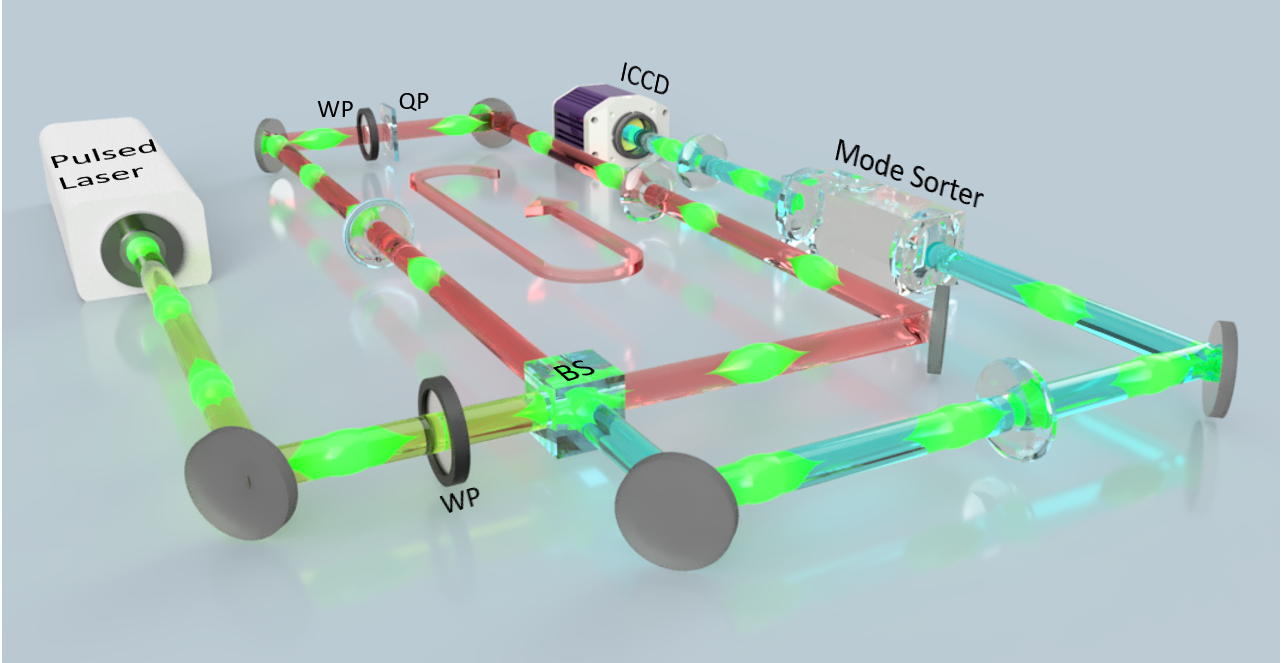}
	\caption{\textbf{Experimental concept.} A bright laser pulse with a Gaussian spatial mode is directed into a looped resonator through a beam-splitter (BS) as shown by the green path. A wave-plate (WP) initializes the state of the walker. The pulse is looped inside the cavity, highlighted by the red path in the indicated direction. Intra-cavity imaging optics to ensure that all spatial modes are divergence compensated. A q-plate (QP) and WP inside the cavity generates the step and coin toss respectively, allowing for the evolution of the walker in OAM space. After each round trip (one step of the walk), a fraction of the light is coupled out of the loop through the BS and imaged, along the blue path, onto the detection system comprising a OAM mode sorter followed by a lens, sorting the modes into transverse positions on an ICCD (Intensified Charge-Coupled Device) gated camera. This concept figure shows many pulses circulating in the resonator, however, experimentally only a single pulse is utilized. Various QWs can be implemented by adjusting the input mode, type of wave-plate and wave-plate (WP) orientation, as detailed in main text.}
	\label{fig:conceptual spatial modeQW}
\end{figure*}

More details are provided in the Supplementary Information. Here the propagator $Z_H$ couples OAM (walker) to the polarisation (coin), i.e. it generates a vector beam from a scalar beam. The non-separability of the polarisation and the spatial degrees of freedom in the vector beam corresponds to the quantum entanglement between the walker and the coin degrees of freedom in realisations of the quantum walk with quantum systems. 

Choosing the polarisation to represent the coin state has the advantage that all coin toss operators $\hat{C}$, which form the group SU(2) of special unitary operators,  can be simply realised by a combination of at most two quarter-wave plates and a half-wave plate \cite{simon1990minimal}. Moreover,  it can be shown that the three parameters (the so-called Euler angles) that characterise any coin toss operator $C\in$ SU(2) can be reduced by local basis transformations on the coin and the walker space to a single parameter that distinguishes the dynamics of all quantum walks with time-independent coin toss operator \cite{goyal2015unitary}. As a result,  any quantum walk can be expressed by a propagator of the form
\begin{equation} 
\hat{Z}_\theta = \hat{C}_{\theta} \hat{C}_N\hat{S}\,,
\label{simpleqw}
\end{equation}
where $ \hat{C}_{\theta}$ has the  matrix representation
\begin{equation}
\hat{C}_{\theta}=\frac{1}{\sqrt{2}}\begin{bmatrix}
\cos \frac{\theta}{2}  & i\sin \frac{\theta}{2}\\
i\sin \frac{\theta}{2} & \cos \frac{\theta}{2}
\end{bmatrix},
\label{eqn:rot}
\end{equation}
\noindent with respect to the basis of circular polarisations $\ket{R}, \ket{L}$. We have inserted a polarisation flip $C_N$ (also known as {\it Not Coin}, see below) to compensate the flip introduced by the q-plate. In this way the coin operators coincide with those conventionally used for quantum walks. The combination $\hat{C}_{\theta} \hat{C}_N$ can be implemented using two quarter-wave plates $\hat{Q}$ and one half-wave plate $\hat{H}$:
\begin{equation}
\hat{C}_{\theta} \hat{C}_N = \hat{Q}_{45^{\circ}} \hat{H}_{\frac{\theta}{4}} \hat{Q}_{45^{\circ}}\,, 
\label{new1}
\end{equation} 
where we have suppressed an irrelevant global phase factor. 

After $n$ steps with propagator $\hat{Z}= \hat{C} \hat{C_N} \hat{S} $ the initial state $\ket{\psi}_{0}$ of the walker will have evolved to the state $\ket{\psi}_{n}$ with
\begin{equation}
\ket{\psi}_{n}= \hat{Z}^{n}\ket{\psi}_{0},
\label{eqn:Quantum walking steps}
\end{equation}

\noindent Our approach makes it possible to study all QW configurations with time-independent coin. By way of example, we implemented the usual QW with a Hadamard coin
\begin{equation}
\hat{C}_{H}=\frac{1}{\sqrt{2}}\begin{bmatrix}
1 & 1\\
1 & -1
\end{bmatrix},
\label{eqn:Hadamard coin}
\end{equation}
\noindent as well as a balanced coin   
\begin{equation}
\hat{C}_{B}=\frac{1}{\sqrt{2}}\begin{bmatrix}
1 & i\\
i & 1
\end{bmatrix},
\label{eqn:Balanced coin}
\end{equation}

\begin{figure*}
\centering
\includegraphics[width=\linewidth]{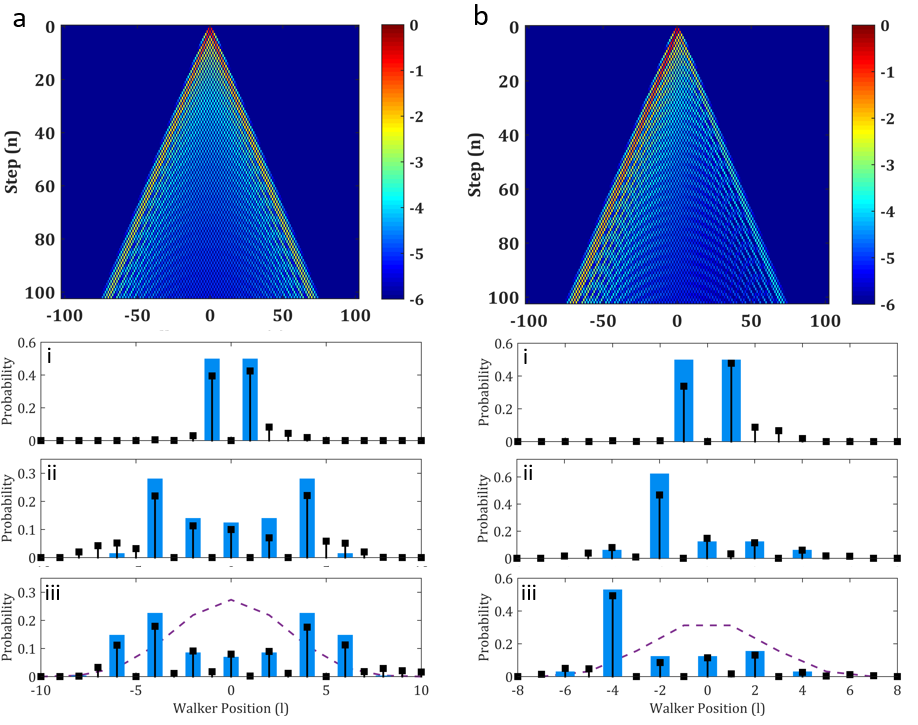}
\caption{\textbf{Experimentally realized Hadamard QWs.} Theoretical density plots of the probability of a QW with a Hadamard coin for 100 steps with a (a) symmetrical and (b) asymmetrical input state (intensity is plotted on a logarithmic scale to emphasize the trend). Sub-plots below the density plots show the experimental realization of the walks for choice steps. Blue bars indicate the simulated probabilities and the black points the experimental data. Respectively, (i) shows Step 1 with S = 0.82 and S = 0.81, (ii) Step 6 with S = 0.80 and Step 4 with S = 0.86 and (iii) Step 8 with S = 0.85 and Step 5 with 0.90. The dashed purple line shows the classical RW distribution for comparison.}
\label{fig:HadQW}
\end{figure*}

\noindent a NOT coin

\begin{equation}
\hat{C}_{N} =\begin{bmatrix}
0				&		1 \\
1   &					0
\end{bmatrix}, 
\label{eqn:NOT coin}
\end{equation}

\noindent and an Identity coin 
\begin{equation}
\hat{C}_{I} = \begin{bmatrix}\
1				&				0 \\
0				  &			    1
\end{bmatrix}.
\label{eqn:Identity coin}
\end{equation}
Demonstrating the versatility of the proposed  setup, each of these coins could be realised with a single wave plate: $\hat{C_H}\hat{C_N}= \hat{Q}_{45^{\circ}}$,  $\hat{C_B}\hat{C_N}= \hat{H}_{90^{\circ}}$, $\hat{C_I}\hat{C_N}= \hat{H}_{0^{\circ}}$, and the NOT coin $\hat{C}_N$ without wave plate since it is already included in the action of the q-plate.

\noindent \textbf{Experiment.}Following the aforementioned theory, we realize a QW using OAM as the position space and polarisation as the coin toss, within a looped resonator cavity that incorporates a $q$-plate, as illustrated in Fig. \ref{fig:conceptual spatial modeQW}, with further details in the Methods and Supplementary Information.  The QW was initiated with a homogeneous vertically polarised ($\ket{V}$) Gaussian beam of pulse duration $\approx 10$ ns, with the initial state determined by the angle of the HWP.  This input state, separable and with 0 OAM, was admitted into the resonator (3 m length) by a beamsplitter and circulated in a 4-f imaging arrangement to mitigate mode divergence.  A percentage of the pulse was then transmitted out of the resonator while the rest was reflected for another round trip.  By placing a $q$-plate (QP) and WP inside the resonator, each round trip corresponds to a step in a QW, with each step in the walker measured by the partial transmission of the pulse after each round trip.  It follows that the QW evolves temporally with each step at multiples of the round trip time. Consequently, as each round trip pulse is partially transmitted by the BS, the entire probability distribution for each position may be simply retrieved with a mode sorter \cite{berkhout2010efficient} and ICCD camera, the latter for determining the step number and the former to detect the walker position in OAM space (see Supplementary Information).  Clearly by restricting the ICCD exposure time, the OAM distribution of the pulse (walker) for a single step may be isolated. The setup thus allows each step to be easily monitored, unlike previous classical light implementation in waveguides \cite{perets2008realization, sansoni2012two} or cascaded q-plates \cite{cardano2017detection}. Moreover, the setup is scalable since the physical resources do not change with the number of steps, but rather remain constant with only one QP-QWP pair required for an entire walk of arbitrary step number. 

\begin{figure}[h]
\centering
\includegraphics[width=\linewidth]{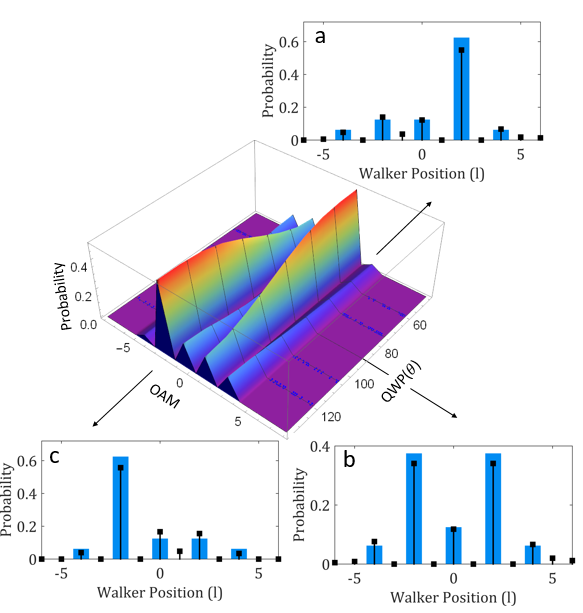}
\caption{\textbf{Rotating the coin.} Theoretical 3D plot of change in QW symmetry with the coin angle with a QWP coin at Step 4. Insets show the theory (blue bars) and experimental data (black points) for a change in symmetry with an asymmetrical input state for the QWP at (a) $45^{o}$ yielding a Hadamard coin ($S=0.92$), (b) at $90^{o}$ yielding a Balanced coin ($S=0.94$) and (c) at $135^{o}$ yielding a full reversal in the Hadamard symmetry ($S=0.94$).}
\label{fig:QWPcoin}
\end{figure}

\begin{figure*}
\centering
\includegraphics[width=\linewidth]{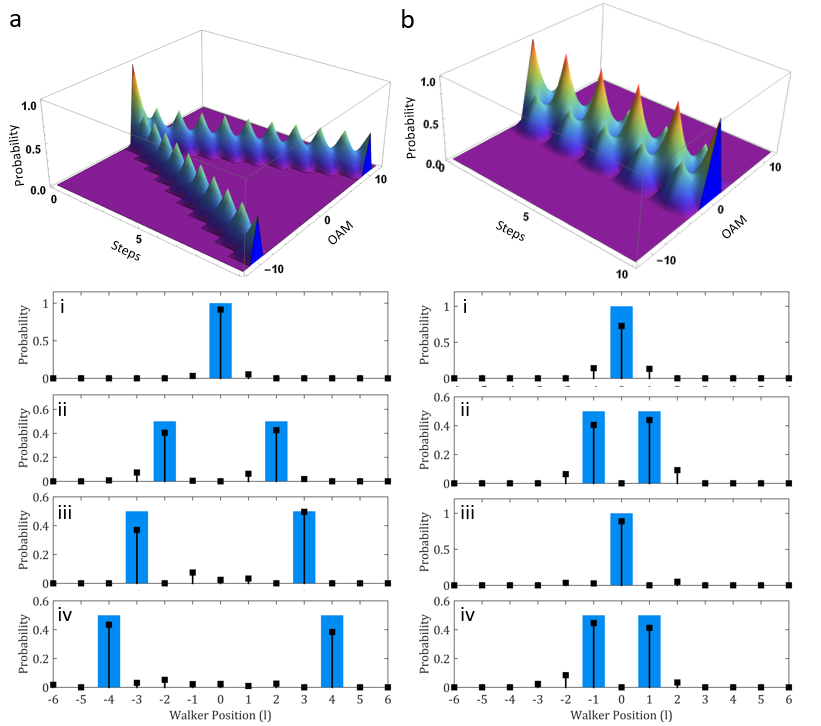}
\caption{\textbf{Experimentally realized QW extremes.} Theoretical 3D plots of the QW probabilities for 10 steps with an (a) Identity coin and (b) NOT-coin. Sub-plots below show the experimental realization of the walks for choice steps. Blue bars indicate the simulated probabilities and the black points the experimental data. Respectively, (i) shows Step 0 with S = 0.92 and S = 0.73, (ii) Step 2 with S = 0.83 and Step 3 with S = 0.84, (iii) Step 3 with S = 0.86 and Step 4 with S = 0.89 and (iv) Step 4 with S = 0.82 and Step 5 with 0.86.}
\label{fig:IDandNOT}
\end{figure*}

\noindent \textbf{Experimental results.} Varying combinations of initial states and coin operators were experimentally realized with the same system and subsequently shown in Figs. \ref{fig:HadQW}, \ref{fig:QWPcoin} and \ref{fig:IDandNOT}. Results for the Hadamard coin operator with symmetrical and asymmetrical initial states are shown in Fig. \ref{fig:HadQW}. This was achieved by placing the QWP with the fast axis at $45^{\circ}$ after the QP while adjusting the fast axis of the HWP before the resonator.

\noindent  For a diagonal input (HWP at $67.5^{\circ}$ acting on a vertically polarised beam) the symmetrical initial coin state evolution is shown in Fig.\ \ref{fig:HadQW} (a). Here a theoretical plot of the QW evolution is given, indicating a distinct and equivalent divergence of the OAM weightings away from the central position ($\ell = 0$) and towards the outer edges. The experimental measured distributions for Steps (i) 1, (ii) 6 and (iii) 8 are consequently shown below, displaying the telltale characteristics of a QW distribution, even though the light is purely classical. Good correlation between the measured and simulated distributions are evident with the lowest calculated Similarity (as defined in the Supplementary Information) being $S =0.80$ .

Similarly, by placing the initializing HWP at $45^{\circ}$, the coin state of the pulse was changed to horizontal polarization, reflecting an asymmetrical input state with respect to the Hadamard coin operator. Accordingly, the QW distribution exhibits asymmetry, as shown in  Fig.~\ref{fig:HadQW} (b), where divergence of the OAM weighting away from the central position is weighted towards one (negative) direction as the walk evolves. Experimental measurements for Steps (i) 1, (ii) 4 and (iii) 5 showcase the experimental actualization of this. Starting with a similar OAM distribution as the symmetrical case at Step 1 ($S = 0.81$), which is centered around the origin, a clearly asymmetric distribution is seen by Step 4 ($S = 0.86$) which is is then further propagated at Step 5  ($S = 0.85$), in accordance with the simulated distribution. Here the asymmetrical nature of the walk is evident with the weightings detected for the OAM values towards the left being approximately 3 times greater than that of the right, alongside a distinct reduction in weightings at the central OAM value.

In both cases, a distinct difference may be seen in comparison to the classical RW distribution for the last measured step, where maximal probability is maintained at the origin with a Gaussian spread outwards.

The relative phase rotations induced by the coin operator action (and thus the waveplate) in the QW, however, also have significant impact on the symmetry of the distribution and thus how the walker propagates. This variation may be clearly seen by changing the orientation of the QWP as it alters the phase rotations between the coin states (right- and left-circular polarization) with the fast-axis angle. The theoretical 3D plot in Fig.~\ref{fig:QWPcoin} illustrates this alteration from the characteristic Hadamard coin operation at $45^{\circ}$, yielding an asymmetrical distribution for a vertically polarized initial state to a completely symmetric distribution at $90^{\circ}$, forming the commonly known Balanced coin. Continued rotation of the angle then results in a reversal of the asymmetry at $135^{o}$ in comparison to the Hadamard case. Experimental results are given for these extreme cases as shown by the insets graphs where good agreement occurs for the (a) Hadamard coin with a similarity of $S=0.92$, (b) Balanced coin with a similarity of $S=0.94$ and (c) a reversed Hadamard coin with similarity $S=0.94$.

As the QW distribution results from the walker interference with itself, it follows that the phenomenon holds extremes in this respect where the walker either does not interfere with itself or fully interferes with itself. These extremes are enacted by the Identity and NOT-coins respectively which are illustrated in Fig.~\ref{fig:IDandNOT} (a) and (b) respectively. By inserting a HWP at $0^{\circ}$ inside the cavity we can produce the identity coin. The 3D plot illustrates the characteristic evolution of this walk over 10 steps. Here the operation of this coin is to continually ladder each state in the same direction, causing no interference between them. As a result, this coin generates states at the extremes of the distribution for each step, such that maximal variance is seen for the QW. Experimental realization of this is shown in plots (i - iv) of Fig.~\ref{fig:IDandNOT} (a), where the only states occupied are the OAM values corresponding to the step number. For instance, (i) Step 0 occupies OAM 0 (S = 0.92), (ii) Step 2 occupies OAM $\pm$2 (S = 0.83) (iii) Step 3 occupies OAM $\pm$3 (S = 0.86) and (iv) Step 4 occupies OAM $\pm$3 (S = 0.82).

The built-in coin resulting from the polarization flipping action of the QP naturally forms the NOT coin, causing the walker to invert polarization states and thus walker direction at each step. The walker then oscillates about the origin, perpetually propagating from $l = 0$ to $\pm1$ and back as shown in the 3D plot of Fig.~\ref{fig:IDandNOT} (b). Results for this coin are shown in Fig.~\ref{fig:IDandNOT} (b) (i-iv). Here the distinct confinement of the distribution to the initial position of the walker is evident throughout the steps. Step 0 (i) (S = 0.73) and Step 4 (iii) (S = 0.89), in particular, shows the walker at the origin, baring significant contrast to the same step distributions shown for Fig.~\ref{fig:IDandNOT}. Steps 3 (ii) (S = 0.84) and 5 (iv) (S = 0.86) highlights the maximum spread possible for this walk with the extreme limited to $\pm1$ for alternate steps.  

\section{Discussion and Conclusion}

Our experiment is based on a resonator type configuration using non-separable states of light, which together allows for multiple steps in a QW to be observed with only a few elements.  In our experiment the step number has been limited by available resources: to fit the detector spectral range the available laser was frequency doubled, resulting in high losses and poor beam quality. Additionally, the q-plate generated OAM states excite unwanted radial modes \cite{karimi2007hypergeometric} and with many passes through the q-plate, significant excitation of the higher order radial modes occu, further contributing to losses \cite{sephton2016revealing}. All these factors can be circumvented in a more engineered version of the experiment.  For this reason our step numbers are modest, but the experiment serves to make the point of scalability and versatility.

Our results support the idea that we may implement a \textit{quantum} walk using \textit{classical} light. It is well known that classical light can be used to mimic a QW, and has been achieved by a complex arrangement of connected interferometers, often on a waveguide \cite{qi2016experimentally} or in a fibre \cite{regensburger2011photon}, akin to a Galton board \cite{bouwmeester1999optical}.  In this way, the quantum principle of superposition is replaced by classical interference of many paths \cite{knight2003quantum}. This approach is experimentally challenging and is not scalable, requiring ever more elements for each increase in step number. 
Our approach, allows the implementation of a QW such that
no additional elements are needed in order to scale the system in the number of steps by exploiting a resonator geometry. We have exploited OAM and polarisation as our non-separable degrees of freedom, the former for the space in which the walker moves. A practical advantage of walking in OAM space relates to the physical size of this "position" space: in traditional classical (and quantum) tests, Galton board arrangements of beam-splitters arrays are required which branches out linearly with the number of steps required. For an OAM position space, all the twists and thus positions are superimposed in the same beam such that a theoretically infinite space for moving the walker exists in a single beam of light.   QWs with in OAM space have been reported using a cascade of optical elements, one after the other, again requiring a scale in elements for an increase in steps \cite{cardano2015quantum}.  The combination of a resonator and an OAM walker space is that (i) it allows any number of steps to be achieved without adjustment to the experimental arrangement; and (ii) all positions in the walker space can be measured at once rather than iteratively.   

Our approach exploits the idea of non-quantum entanglement generation in an optical resonator, with the resulting non-separability equivalent to a quantum implementation albeit with classical
light.  We predict that this holds the key for future advances in QWs: there is no equivalent to the "no-cloning" theorem in classical optics, implying that the beam losses inside the resonator can be compensated for with a suitable gain medium.  If the gain can be made equal for all modes in the non-separable superposition, which we acknowledge is a challenging task and the topic of current research \cite{sroor2018purity}, then the QW can be run for virtually unlimited steps with just a few elements, eventually limited in step number by the aperture size of the optical arrangement.  Such an arrangement $-$ a resonator that amplifies light at each round trip, leaking a little out for measurement $-$ would in essence be a \textit{quantum walk laser}.  

In conclusion, a central message of this work is to refute the mantra of attenuating laser light, ostensibly to reach the single photon level, to make the QWs truly quantum.  We believe that this work is the first to make explicitly clear, by way of experiment, that quantum walks with a single walker do not require quantum states of light.

In conclusion, a central message of this work is to refute the mantra of attenuating laser light, ostensibly to reach the single photon level, to make the QWs truly quantum.  We believe that this work is the first to make explicitly clear, by way of experiment, that quantum walks with a single walker do not require quantum states of light.
Further, by employing a resonator type configuration we are able to make the system scalable without additional resources, in contrast to other approaches using classical and quantum light.  Finally, we outline how it may be possible to increase the number of steps dramatically by introducing the concept of a quantum walk laser.  We believe that our approach offers a new path for future QW implementations that are both versatile and scalable.

\section*{Materials and correspondence}
Correspondence and requests for materials should be addressed to AF.


\section*{Acknowledgments}
Andrew Forbes would like to that the NRF-CSIR Rental Pool Programme for equipment use.  The authors would like to thank Darryl Naidoo for useful discussions as well as Ameeth Sharma and Gary King for technical assistance.


\section*{Authors' contributions}
BS performed the experiment and analysis of the data under supervision of AD and AF, while SG, FR and TK provided theoretical input.  FR, LM and MP provided custom designed optics for the experiment. All authors contributed to the the data interpretation and writing of the manuscript. AF supervised the project.



\section*{Competing financial interests}
The authors declare no financial competing interests.

\clearpage
\newpage
\section{Supplementary Information}

\subsection{Traditional quantum walk}
\begin{figure}
	\centering
	\includegraphics[width=\linewidth]{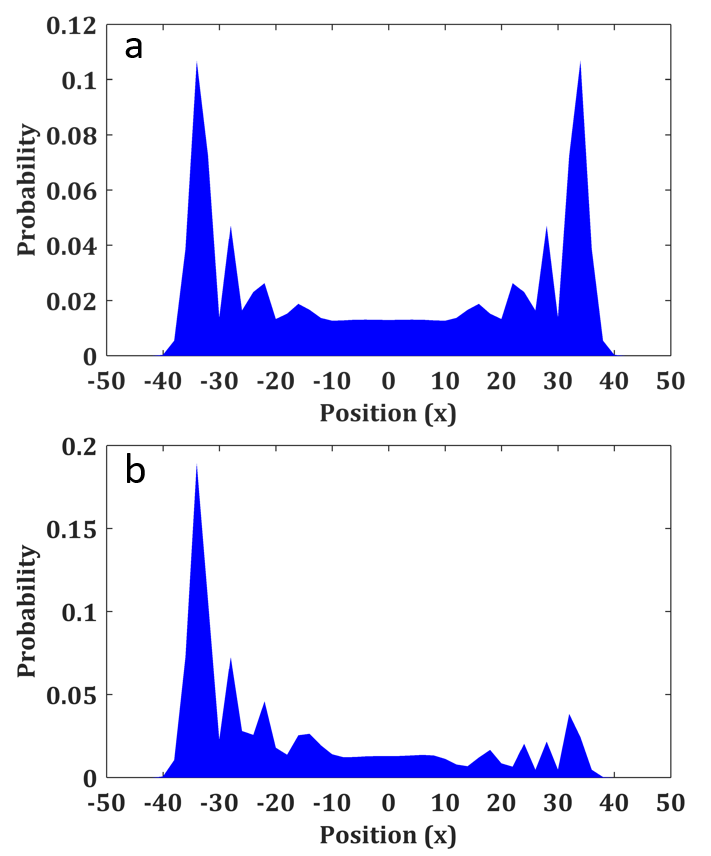}
	\caption[QW Probabilities]{Graph of the probability distribution for a 1-dimensional Hadamard QW after 100 successive movements or steps for a (a) symmetrical and (b) asymmetrical input state.}
	\label{fig:QWProbDist}
\end{figure}

Discrete time quantum walks evolve in the Hilbert space,  ($\mathcal{H}_{x} \otimes  \mathcal{H}_{c}$) of position and coin subspaces, respectively, whereby each position ($\{\ket{x}\}^{\infty}_{-\infty}$) occupied by the walker has an internal coin state ($\{\ket{H},\ket{T}\}={\ket{c}}^{T}_{c\: =\: H}$) associated with it. Evolution of the walker is then achieved by successive unitary operations of a coin operator such as the Hadmard coin,
\begin{equation}
\hat{C}_{H} = \frac{1}{\sqrt{2}} \left( \ket{H} + \ket{T})\bra{H} + (\ket{H} - \ket{T} \right) \bra{T},
\label{eqn:HadamardCoin}
\end{equation}
acting on the walker internal coin state at each position, flipping the coin and a shift operator, 
\begin{equation}
\hat{S} = \sum^{\infty}_{x\: =\: -\infty} \ket{x+1}\bra{x} \otimes \ket{H}\bra{H} + \sum^{\infty}_{x\: =\: -\infty} \ket{x-1}\bra{x} \otimes \ket{T}\bra{T},
\label{eqn:ShiftOperator}
\end{equation}
that propagates the walker right (left) at each position according the internal heads (tails) coin state. Concatenation of these operations to generate the step operator for the example of a Hadamard walk,
\begin{eqnarray}
\hat{U}_{H} = \hat{C}_{H} \hat{S} = \sum_{x\: =\: \infty}^{-\infty} \left( \ket{x+1}\bra{x}\otimes\frac{1}{\sqrt{2}}(\ket{H}+\ket{T} \right) \bra{H} \\  \nonumber
+ \sum_{x\: =\: \infty}^{-\infty} \left( \ket{x-1}\bra{x}\otimes\frac{1}{\sqrt{2}}(\ket{H}-\ket{T} \right) \bra{T}, 
\label{eqn:QWalegra}
\end{eqnarray}
then results in one full step when applied to the walker with an initial state such as $\ket{\psi}_{0} = \ket{0} \otimes \ket{H}$. Implementation of the step operator, causes the state of the walker to change according to the number of steps or implementations, $n$, 
\begin{eqnarray}
\ket{\psi}_{n} = \hat{U}_{H}^{n}\ket{\psi}_{0} = \sum_{x\: =\: 0}^{d}[c_{H,x}\ket{H,x} + c_{T,-x}\ket{T,-x}]
\label{eqn:walkingAlgebra}
\end{eqnarray}
where $c_{H,x}$ and $c_{T,-x}$ are complex amplitudes indicating the probability of the walker occupying each position and $d = (2n+1)$ is the dimension of the space occupied by the walker. With each step, the walker moves to adjacent positions on the 1D line. Subsequently, when occupying two consecutive positions before the step, overlap in positional occupation occurs for shared movements. The complex amplitude of the walker thus interferes to generate a different probability distribution over the position spaces than the classical random walk \cite{kempe2003quantum}. Measurement of the QW superposition collapses the superposition, forcing the walker to localize at a particular position ($x$) with the associated probability
\begin{eqnarray}
P_{x,n} = |\braket{x|\psi_{n}}|^{2}.
\label{eqn:Meas}
\end{eqnarray}
Figure \ref{fig:QWProbDist} shows the probability distribution $P(x)$  for the walker after taking $n = 100$ steps for symmetrical and asymmetrical initial states with a Hadamard coin. 
Here it can be seen that the probability of finding the walker is closest to the ends of the distribution as the walker destructively interferes with itself at the center. Moreover, by changing the phase of the initial state, the interference may also be altered to generate a distribution where the greatest probability is weighted more to one direction of the position space as seen by the larger spike in probability to the left in (b).

Action of the coin operator additionally causes the coin and position states to become entangled as may be seen in the non-separable form of Eq. \ref{eqn:walkingAlgebra}. It is these dynamics which causes the QW to obtain the different characteristics such that it may be exploited for the simulation and computation applications with up to ballistic speedups.
\subsection{Detailed experimental setup}
\begin{figure*}[t]
	\centering
	\includegraphics[width=\linewidth]{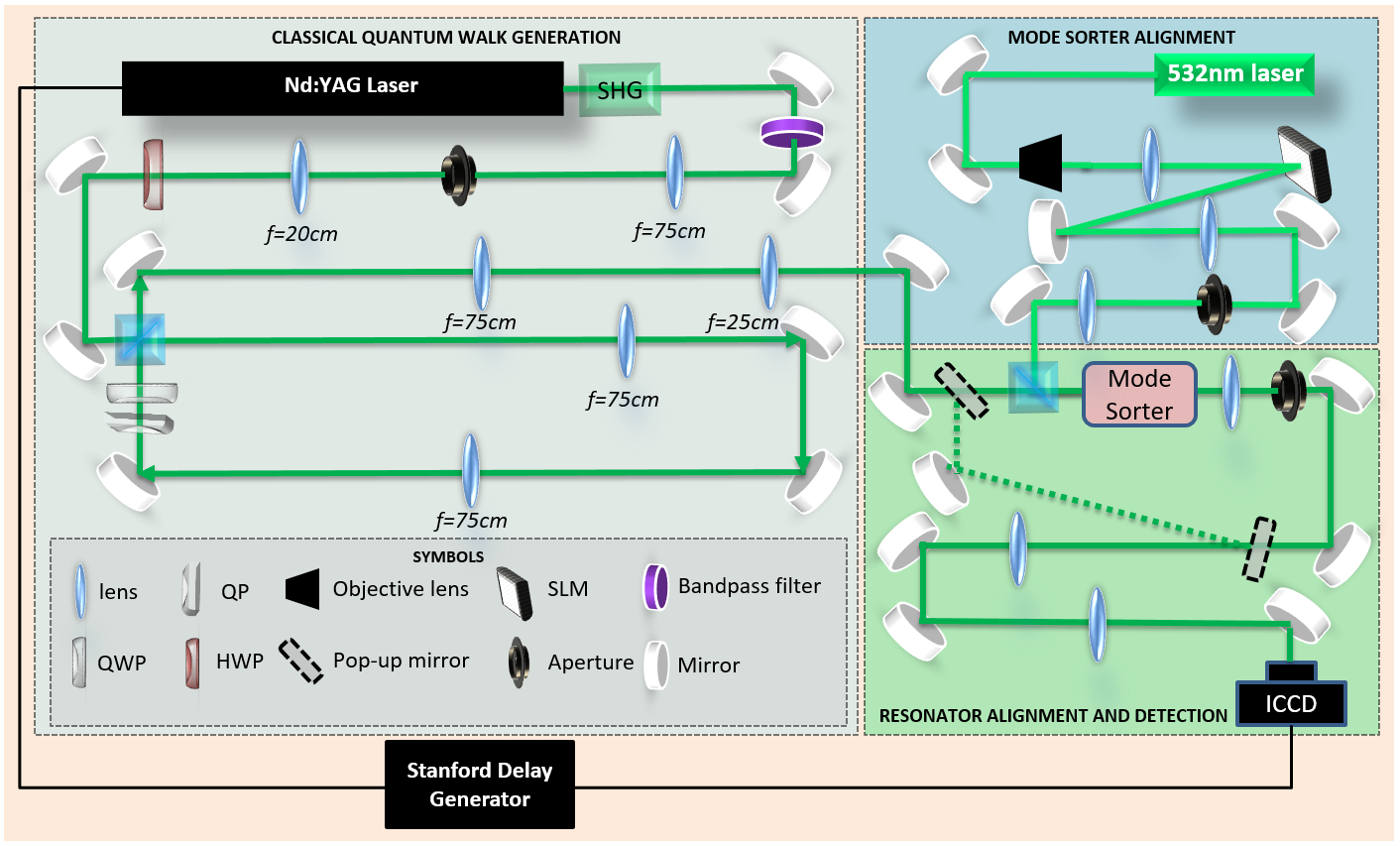}
	\caption[The setup for a spatial mode QW]{Schematic of the actual setup constructed for physical implementation of the QW.}
	\label{fig:QW setup}
\end{figure*}
A schematic of the actual experimental setup is given as Fig. \ref{fig:QW setup}. Here a pulsed laser (Spectraphysics Quanta-Ray DCR-11) at a wavelength of 1064 nm generated a single input light pulse with 0 OAM (initial position of the walker).  As we did not have an  gated detector for this wavelength the laser was frequency adjusted by second harmonic generation, resulting in the high losses of the system and hence limiting the maximum number of steps possible.  To this end a non-linear SHG crystal converted the wavelength to 532 nm through frequency doubling to remain within the ICCD (iStar AndOR) detection range. This step prohibited amplification within the cavity due to the lack of suitable gain media at this wavelength.  Structuring a Gaussian intensity distribution of appropriate beam size was accomplished through a spatial filter which led to significant losses, thus limiting the maximum step number we could achieve. Propagation of the pulse through a HWP served to prepare the input polarization state symmetry (e.g., diagonal polarization for a symmetrical Hadamard QW), after which, it was injected into the resonator (3 m perimeter) by a 50:50 non-polarizing beam-splitter. Placement of the QP ($q = 0.5$) and WP concatenation initialized the QW and advanced it by one step with each consecutive round trip. The output pulse from the beam-splitter was subsequently imaged from the QP plane to the mode sorter.   Alignment of the mode sorter was attained by constructing an adjoining OAM mode generation setup (see Supplementary Information). Here a 532 nm wavelength diode laser was expanded with a $10\times$ objective lens and collimated through an $f=300$ mm lens onto an SLM where phase and amplitude modulation was utilized to generate superpositions of LG beams.  A 4-f system was built to isolate and image the 1st diffraction order onto the mode sorter. The SLM generated mode was then combined and aligned with the output mode from the resonator with a 50:50 BS. By passing test OAM superpositions from the SLM through the MS, optimal alignment of the elements was then achieved. The Fourier plane of the MS configuration was directed and imaged onto the ICCD plane with another 4-f system. Choice placement of a popup mirror before the MS and within the FP imaging system allowed the QP plane image to be re-imaged onto the ICCD with a lens placed between the pop-up mirrors. The second lens in the MS imaging system then served a dual purpose as the second imaging lens in the 4-f re-imaging system of the QP plane. This allowed the output beam structure for each pulse to be individually captured for each round trip, simplifying the alignment process.  Subsequent capture of each round-trip pulse was achieved through utilization of an iStar AndOR ICCD camera with a temporal resolution on the 10 ns scale. Synchronization between the initial laser pulse and recording window of the camera was attained with a Stanford delay generation working in combination with the iStar on-board digital delay generator.

\subsection{Action of the q-plate}
The $q$-plate (QP) used in our experiment was a patterned liquid crystal static element which imparted geometrical phase to the light field transmitted through. Here the charge ($q$) of the plate dictates the OAM value generated while the polarization of the incoming beam controls the handedness of the OAM generated \cite{marrucci2006optical}. It follows that polarization could thus be used as a control for the laddering of OAM to higher or lower values in either handedness (positive or negative OAM). Operation of the QP in the circular polarisation basis is given by the Jones matrix \cite{marrucci2012spin, piccirillo2013orbital}
\begin{equation}
QP = \begin{bmatrix}
0		&		ie^{-i2q\phi}	 \\
ie^{i2q\phi}		&		0
\end{bmatrix}
\label{eqn:QP matrix CP}
\end{equation}
Incident right circular polarisation (RCP), $\ket{R} = [1;0]$, then gains the phase, $e^{i2q\phi}$, resulting in an OAM of $l = 2q\hbar$ per photon. The polarization is subsequently flipped to left circular polarisation (LCP) in the process. Similarly, LCP, $\ket{L} = [0;1]$, acquires the phase $e^{-i2q\phi}$, resulting in negative OAM of $l = -2q\hbar$. The $i$-value multiplying the phase terms is global and thus may be ignored. It follows that the QP operation may be condensed into the following selection rules:
\begin{subequations}
\begin{align}
\begin{split}
\hat{QP}\ket{l,R} = \ket{l+2q, L}
\end{split} \\
\begin{split}
\hat{QP}\ket{l,L} = \ket{l-2q, R}
\end{split}
\end{align}
\label{eqn: QP selection rules}
\end{subequations}

Additionally, this effective twisting of the light beam produced by geometric phase has further implications in the physical interpretation whereby the CP polarization may also be seen in terms of spin angular momentum (SAM). Here when RCP is incident on the QP, OAM of $2q\hbar$ per photon is generated, and the flip in CP corresponds to a flip in SAM from $1\hbar$ per photon to $-1\hbar$. It is well known that transference of SAM and OAM can occur between light and certain matter \cite{piccirillo2013orbital}. Here, SAM interaction occurs in optically anisotropic media such as birefringent material and OAM in transparent inhomogeneous, isotropic media \cite{marrucci2006optical}. The combination of a thin birefringent (liquid crystal) plate with an inhomogeneous optical axis in the QP subsequently results in the element coupling these two form of angular momentum such that flipping in the SAM may be seen to generate OAM, making the QP a spin-to-orbital angular momentum converter (STOC) where the symmetry of the optical axis patterning effects the conversion values \cite{piccirillo2013orbital}.

\begin{figure}
\centering
\includegraphics[width=\linewidth]{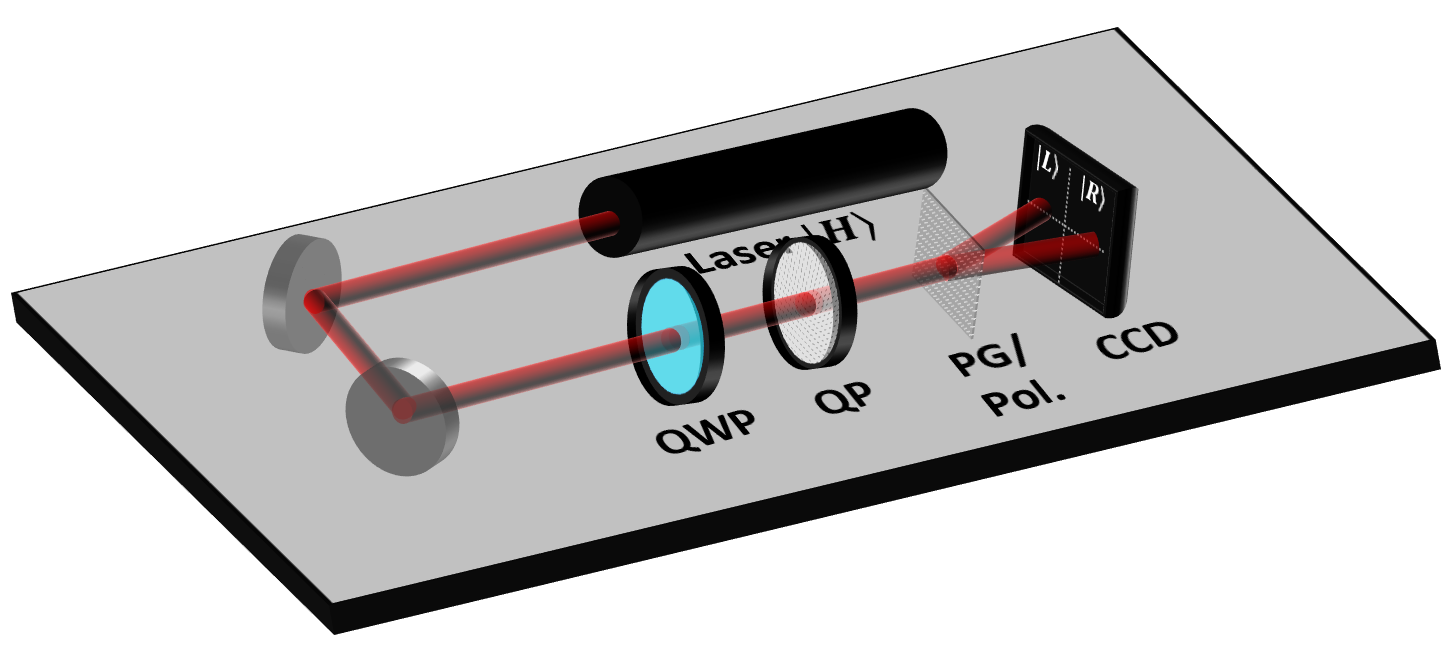}
\caption[SAM QP test setup]{(a) Schematic of the experimental setup used to examine the SAM conversion and mode generation of a $q=0.5$ QP.}
\label{fig:QPsamtestexp}
\end{figure}

Characterisation of the $q = 0.5$ QP used was then carried out. Figure \ref{fig:QPsamtestexp} illustrates the experimental setup implemented to achieve this. A horizontally polarized HeNe laser (wavelength 633 nm) was shone through a QWP before being incident on the QP. A polarization grating (PG) was placed before a Spiricon SPU620 camera which acted to spatially separate the left and right CP of the QP generated beam. 

\begin{figure}[!]
\centering
\includegraphics[width=\linewidth]{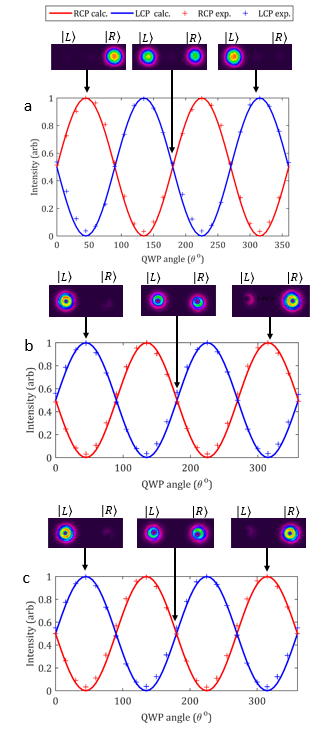}
\caption[Experimental results for SAM conversion by QP]{Spin angular momentum transformations by the QP for varying incident superpositions as shown in (a) with left and right CPs. Here (b) are the corresponding output modes when incident upon the front side of the QP as well as (c) the back side of the QP.}
\label{fig:QP SAM results}
\end{figure}

Variation of the incoming SAM or CP onto the QP was obtained by rotating the QWP fast axis. It follows that superpositions of SAM with various weightings was incident onto the QP based on the QWP angle. These input weightings are shown in Fig \ref{fig:QP SAM results} (a) through projection into RCP and LCP states. The calculated and measured outcome of passing these CP superpositions through the QP is given in Fig. \ref{fig:QP SAM results} (b) which was also projected into the CP basis. Comparison of these two figures show that the LCP and RCP input weightings are inverted after passing through the QP. For instance, at $45^{o}$, RCP generated by the QWP is detected as LCP after passing through the QP. Similarly, at $135^{o}$, the generated LCP is converted to RCP after the QP. At $105^{o}$, CP state with majority weighting is changed from LCP after the QWP to RCP after the QP. It follows that the QP acts to invert the SAM of the incident beam. 

Further observation of the spatial modes of the beams can be seen from the insets. Here the Gaussian profile of the input beam is evident in Fig. \ref{fig:QP SAM results} (a) with the false colour map. The spatial profile of the beam after the QP shows the doughnut distribution with a central intensity null, characteristic of OAM carrying beams. These modes consequently, indicate that OAM is generated by the QP along with the reversal of CP for both incoming LCP and RCP as well as superpositions thereof, as expected from the selection rules in Eq. \ref{eqn: QP selection rules}.  

Moreover, by altering the face through which the beam was incident from front to back, the directional consistency of the element was determined. The experimental outcome is given in Fig \ref{fig:QP SAM results} (c) where the QP side of incidence was reversed in the setup, causing the incoming beam to traverse through the 'back' of the element. Comparison of Fig \ref{fig:QP SAM results} (b) and (c) shows the experimentally measured projections are identical with the QP reversed, enacting the same SAM inversion on the incoming beam. Therefore, it may be concluded that the QP operation follows a directional invariance in performance.

\subsection{A classical entanglement generator}

When the input to the QP is a linear polarisation state, say horizontal, then the output may be expressed as 
\begin{equation}
\hat{QP}\ket{0,H} = \frac{i}{\sqrt{2}}\hat{QP}[\ket{0,L}-\ket{0,R}] = \frac{i}{\sqrt{2}}[\ket{-1,R}-\ket{1,L}]
\label{eqn: applying QP to V}
\end{equation}

\begin{figure}[b]
\centering
\includegraphics[width=\linewidth]{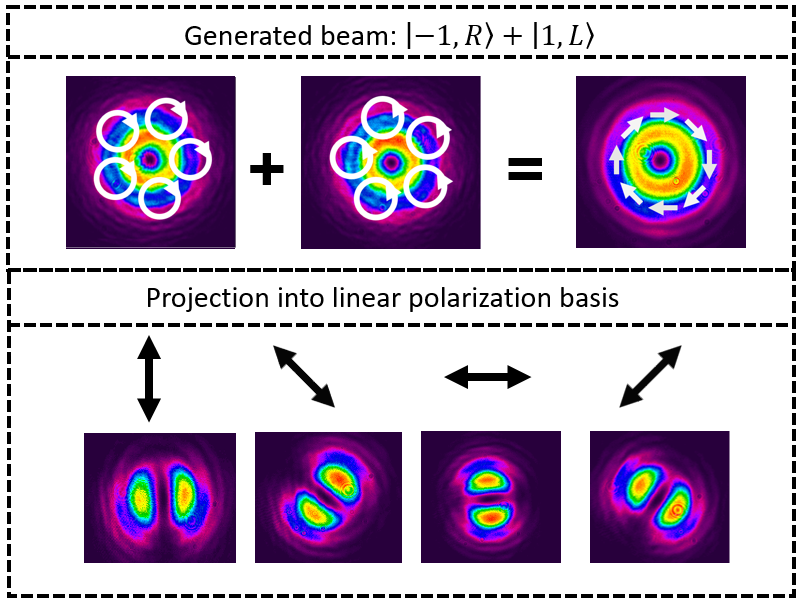}
\caption[The QP classically entangling light]{Example of the QP classically entangling polarization and spatial degrees of freedom with the generation of an azimuthally polarized vector vortex beam from a vertically polarized Gaussian input (scalar beam).}
\label{fig:QP entanglement}
\end{figure}

From the expected states formed, the spatial mode described by OAM = $-1$ is paired to RCP while OAM = 1 is paired to LCP. As a result, these OAM and polarization degrees of freedom in the beam form a non-separable relation such that neither can be factored out. Experimentally, the generated mode is shown as the rightmost upper beam in Fig. \ref{fig:QP entanglement} with a doughnut spatial distribution. Separation of this mode into the CP basis yielded the modes to the left of the image where the respective CPs are superimposed on the images. 

By replacing the PG shown in the experimental setup of Fig \ref{fig:QPsamtestexp} with a linear polarizer, the generated mode was projected into the linear polarisation basis. Here, the non-separability of the spatial mode can be seen where lobes are detected that rotate with the polarizer orientation. Specifically, from the leftmost projected distribution, vertical orientation of the lobes coincides with the vertical orientation of the polarizer. Rotating the polariser though to anti-diagonal, horizontal and diagonal orientations as depicted by the arrows above these images shows the lobes assuming the same directionality. It follows that the projective measurements yield the polarization distribution overlaid on the image of the generated beam. The subsequent pairing of the OAM and SAM modes in this instance resulted in an azimuthal vector vortex beam being generated. Accordingly, the QP may be seen as a "classical entanglement" generator such that orthogonal modes may be both paired and laddered with this element. 

\subsection{OAM detection}
\begin{figure}
\centering
\includegraphics[width=\linewidth]{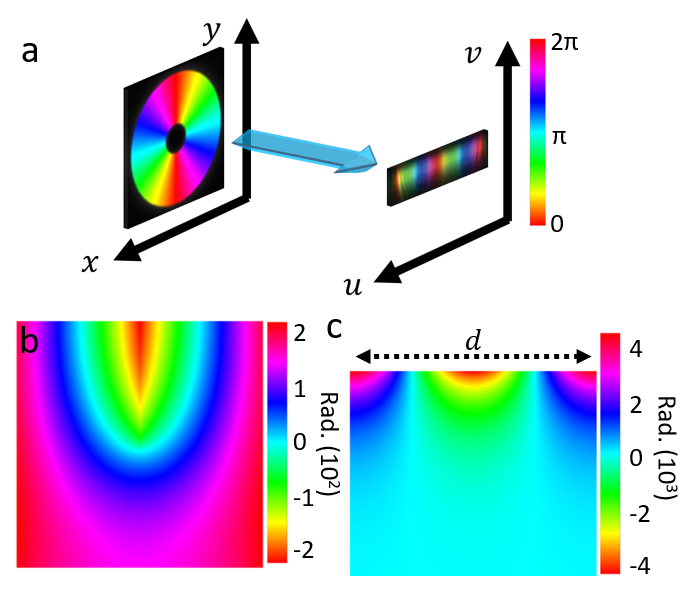}
\caption[Mode sorting with geometrical phase transformation]{The (a) illustration of a conformal mapping which ‘unwraps’ an OAM = 2 mode to a transverse phase gradient and the phase distributions preforming the (a) mapping transformation and (b) correction due to path length differences.}
\label{fig:ModeSorting GeoPhase}
\end{figure}

 Orbital angular momentum mode sorting relies on the application of geometric coordinate transformation. The technique takes advantage of the circular geometry associated with OAM such that a geometrical mapping translates circular to rectangular geometry \cite{berkhout2010efficient} as illustrated in Fig. \ref{fig:ModeSorting GeoPhase} (a). The resultant phase distribution unwrapping causes OAM to be transformed into transverse momentum with a linear phase gradient \cite{mirhosseini2013efficient} as demonstrated in the $(u,v)$ coordinate space of the figure.

Physically, this $(x,y) \rightarrow (u,v)$ transformation is achievable through application of a phase distribution, described in Eq. \ref{eqn: MSelement1}.
\begin{equation}
\varphi_{1}(x,y) = \frac{d}{\lambda f}[y \tan^{-1}(\frac{y}{x})-x\ln(\frac{\sqrt{x^{2}+y^{2}}}{b})+x]
\label{eqn: MSelement1}
\end{equation}
Visualization of the distribution is shown in Fig. \ref{fig:ModeSorting GeoPhase} (b). Here $d$ is the fixed unwrapped beam length, $b$ affects the location in the $(u,v)$ plane, $\lambda$ is wavelength and $f$ the transforming lens focal length. Associated phase distortions in the ‘unwrapped’ beam from optical path length variation requires correction by a second phase distribution, described in Eq. \ref{eqn: MSelement2} \cite{berkhout2010efficient, mirhosseini2013efficient} and Fig \ref{fig:ModeSorting GeoPhase} (c).  

\begin{equation}
\varphi_{2}(u,v) = \frac{db}{\lambda f} e^{\frac{-2\pi u}{d}}\cos(\frac{2 \pi v}{d})
\label{eqn: MSelement2}
\end{equation}

\begin{figure}
\centering
\includegraphics[width=\linewidth]{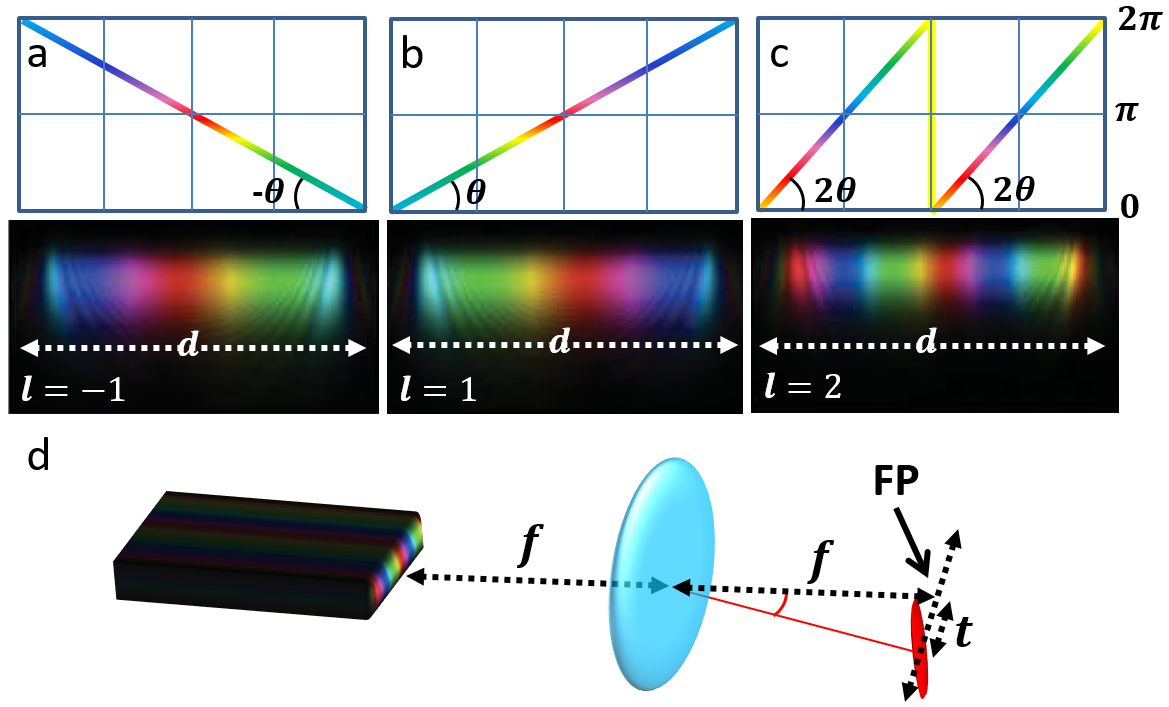}
\caption[Unwrapping and sorting OAM]{Colour map illustration of the phase gradient resulting from the OAM geometric transformation for (a) $l$ = -1, (b) $l$ = 1 and (c) $l$ = 2. (d) Depiction of the sorting action performed by a Fourier transforming lens after the phase correction element.}
\label{fig:ModeSorting}
\end{figure}

The resultant phase distribution unwrapping causes OAM to transform to transverse momentum as it propagates. A transverse phase gradient of $e^{il \tan^{-1}{(\frac{y}{x})}} = e^{il \frac{2 \pi v}{d}}$ across the beam length is then generated \cite{berkhout2010efficient, mirhosseini2013efficient, lavery2013efficient}. The unwrapped beam in the Fourier plane (FP) of a lens forms a diffraction-limited elongated spot \cite{mirhosseini2013efficient}. As the unnwrapped mode contains a phase gradient limited to the length, $d$, each OAM mode results in a transverse phase gradient that is integer-multiples of the other (shown in Fig. \ref{fig:ModeSorting} (a)). 

The lens then forms the spot at a gradient related position in the focal plane due to its Fourier transforming action. The spot position ($t$) is then OAM-dependent \cite{berkhout2010efficient}.
\begin{equation}
t = \frac{\lambda f l}{d}
\label{eqn: spotPOS}
\end{equation}
Moreover, intensity of the spot indicates the ‘amount’ of any OAM-mode present. The mode sorter technique employed with refractive elements allows for efficient detection of a large range of OAM modes and the associated weightings, enabling detection of low intensity sources and in comparison to other techniques such as the SLM modal decomposition mentioned earlier.

A distinct disadvantage occurs with this technique, however, when considering cross-talk between adjacent modes. Due to the finite unwrapped beam size, the width of the spot created in the FP is diffraction-limited, resulting in a constant overlap between modes \cite{berkhout2010efficient, mirhosseini2013efficient}. It was subsequently demonstrated that a maximum of 80\% may be achieved in the required position with good alignment, while the other 20\% spreads into the adjacent mode positions \cite{lavery2013efficient}. Increasing the unwrapped beam size may appear to remedy the situation as it would increase the separation distance between the spots. This, however, also decreases the phase gradient, resulting in the spot width to spacing ratio remaining similar \cite{mirhosseini2013efficient}. Consequently, the overlap between modes is an intrinsic property of the system. A solution was suggested by Berkhout et al. \cite{berkhout2010efficient}, though, whereby the unwrapped beam length may be increased through simply copying the unwrapped modes and placing them next to each other.  

As the linear gradient ranges in periods of 0 to $2\pi$ for every OAM value, placing copies alongside the other does not disrupt the unwrapped beam configuration, allowing this technique to decrease the spot width without altering the spacing \cite{mirhosseini2013efficient,o2012near}. The copied beams then add an extra $N_c$ rotations to this 0 to $2\pi$ periodic gradient.

Further investigation and subsequent implementation was then carried out by including two additional phase transformations on SLMs after the mode sorter optics \cite{mirhosseini2013efficient,o2012near}. Here the first transformation copied the phase-correction unwrapped beam according to the phase term:

\begin{equation}
\varphi_{FOE}(x) = \tan^{-1} \left[ \frac{\sum^{N}_{m = -N} \gamma_{m} \sin((\frac{2\pi \omega}{\lambda})mx + \alpha_{m})}{\sum^{N}_{m = -N} \gamma_{m} \cos((\frac{2\pi \omega}{\lambda})mx + \alpha_{m})} \right]
\label{eqn: fan out elements}
\end{equation}

which resulted in $N_c = 2N+1$ copies placed alongside each other based on the angular separation, $\omega$, in the $x$-direction, causing the element to be labeled a fan-out element (FOE) \cite{mirhosseini2013efficient}. $\gamma_m$ and $\alpha_m$ serve as phase and intensity parameters, respectively, for the different orders ($m$). The second transformation subsequently served as a correction for this fan-out operation.

Ruffato et al. recently combined the log-polar coordinate and fan-out functions as well as the respective phase corrections to condense the operations into two diffractive elements \cite{ruffato2017test}. Here the copying element was combined with the unwrapper so that the beam could be simultaneously copied $N_c$ times and unwrapped alongside each other before encountering the second element \cite{ruffato2018electrically}. Subsequently, the correction terms for the path length difference from the unwrapper is combined with the corrections for joining the copied beams such that a dual and once-off correction is carried out on the beam. It follows that more accuracy should be expected with this method in addition to the convenience of utilizing a more compact system, where the alignment for the system is restricted to half the elements given in the scheme implemented by Mirhosseini et al. \cite{mirhosseini2013efficient}.

\begin{figure}[h!]
\centering
\includegraphics[width=\linewidth]{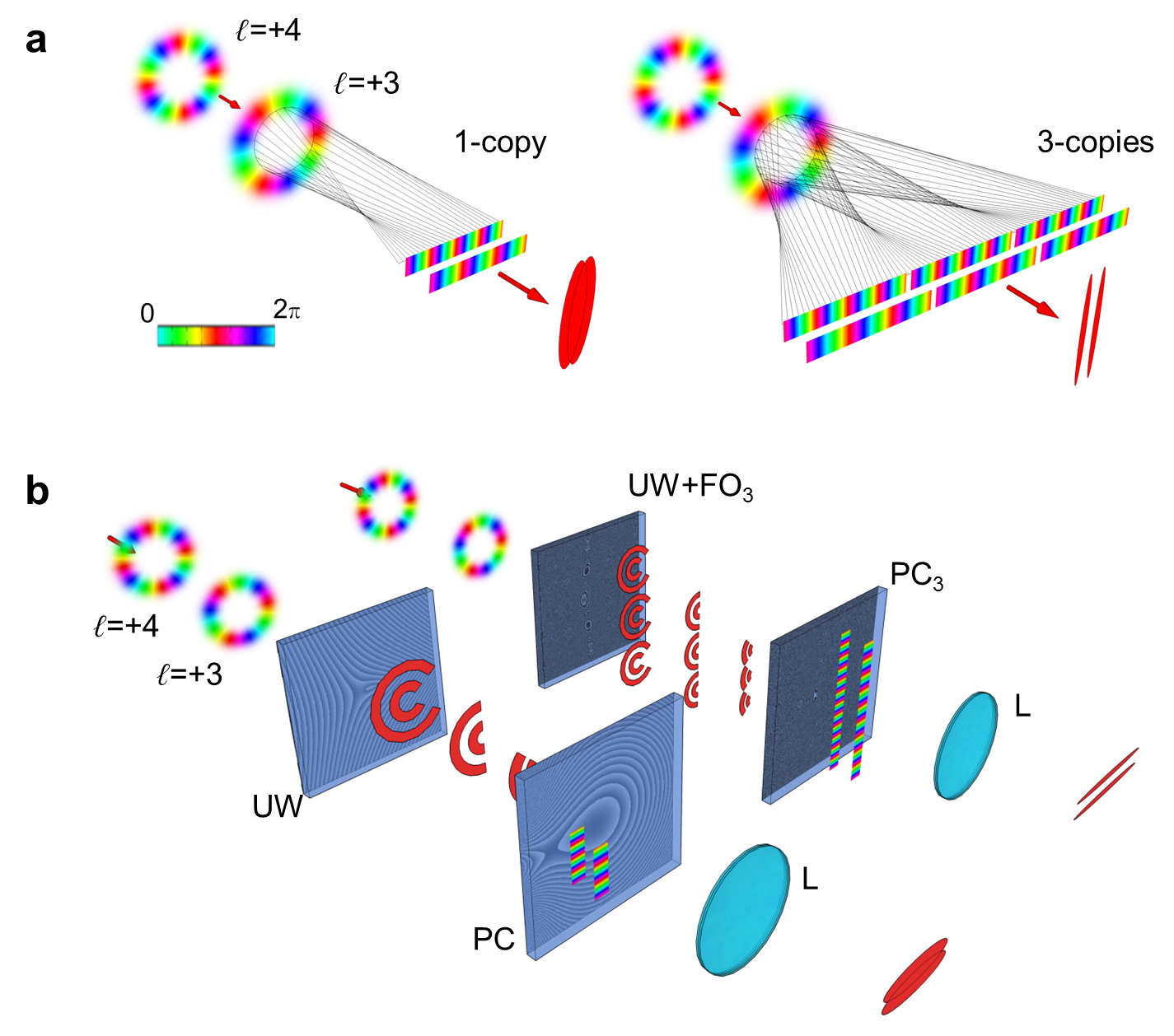}
\caption[Diffractive mode sorters]{(a) OAM demultiplexing with log-pol optical transformation without and with the integration of a 3-copy optical fan-out. The generation of multiple copies increases the spatial extent of the linear phase gradient and therefore reduces the width of the final spots after Fourier transform with a lens. (b) Sorting of OAM beams with log-pol optical transformation in the traditional architecture with the two optical elements in sequence: unwrapper (UW) and phase-corrector (PC). The integration of a fan-out term producing multiple copies (fan-out unwrapper UW+FO3, double phase-corrector PC3) increases the OAM resolution.}
\label{fig:DiffMS}
\end{figure}

The subsequent working principal behind the elements by Ruffato et. al. is depicted in Fig. \ref{fig:DiffMS}.

\subsubsection{Mode sorter performance}

To characterize the expected performance in detecting the QW, analysis of both the refractive and diffractive mode sorting elements used in the experiement was carried out. Table \ref{Table:ModeSortingParameters} gives the fabrication parameters of the tested sorters.

\begin{table}[h]
\centering
\caption[Mode sorter design parameters]{\bf{Comparison of design parameters for the refractive and diffractive sorters.}}
\label{Table:ModeSortingParameters}
\begin{tabular}{m{0.45\columnwidth}>{\centering\arraybackslash}m{0.15\columnwidth}>{\centering\arraybackslash}m{0.23\columnwidth}}
\toprule
\textbf{Design Parameters} & \textbf{Refractive Sorter} & \textbf{Diffractive Sorter}\\
\hline
Element diameter (mm) & 12.7 & 2.0 \\
\hline
Incident beam radius (mm) & $\leq$ 2.10 & 0.500--0.800 \\
\hline
Design wavelength (nm) & 633 & 632.8 \\
\hline
Wavelength range, $\lambda$ (nm) & 400--1000 & 532--732 \\
\hline
Separation between elements (mm) & 300.5 & 8.500 \\
\hline
Unwrapped beam length, $d$ (mm) & 10.5 & 1.12 (1--copy) \\
& & 0.500 (3--copy) \\
\hline
\end{tabular}
\end{table}

The tested diffractive sorters varied in the number of copies that were created where 1 and 3 copies were generated respectively. As the 1-copy and refractive sorter both generated a single unwrapped beam, the expected difference in performance between the sorters was limited to the range of OAM modes that could be accurately sorted. This is due to the refractive sorter being able to receive beams of lager radii, thus compensating for the increased deviation of the angle of incidence for the beam rays directed to the sorter as the OAM increases. A higher range of the sorted OAM values being viably accurate \cite{lavery2013efficient} should thus occur. As a result, evaluation of the refractive sorter was restricted to the comparative OAM ranges while the 1- and 3-copy sorters were further characterized, allowing a more accurate comparison due to the parameter similarities.

\begin{figure}[b]
\centering
\includegraphics[width=\linewidth]{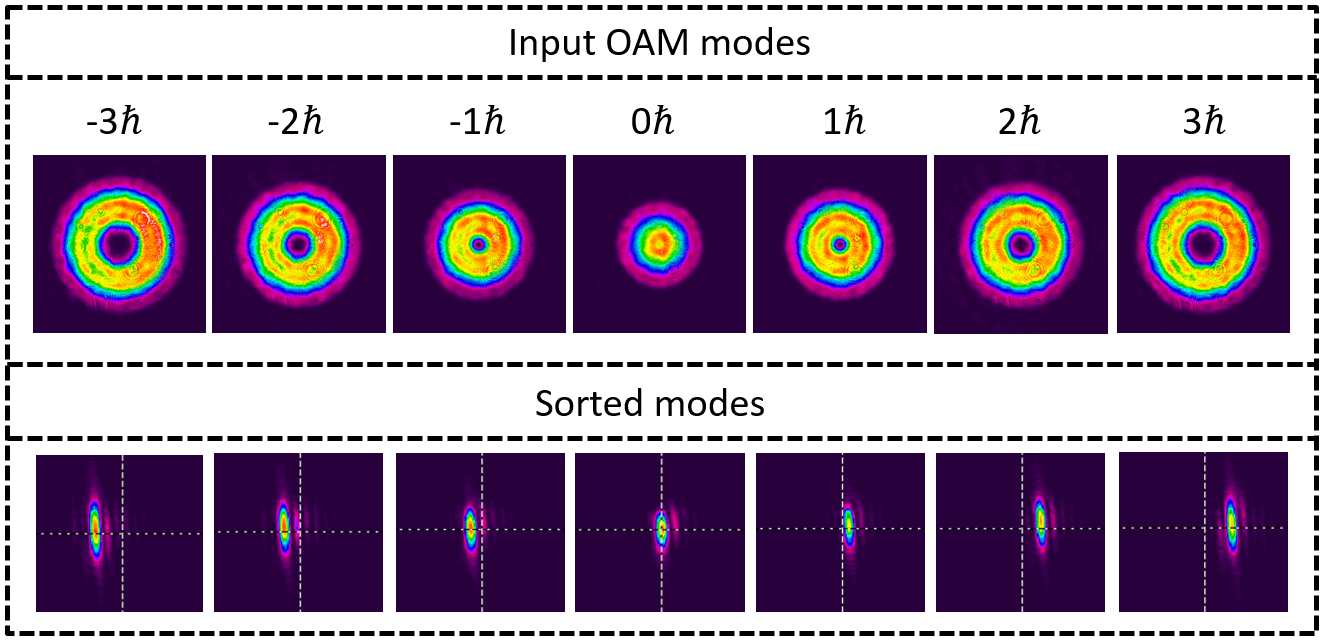}
\caption[Experimentally sorted modes]{Illustration of the sorting action of a MS (bottom row) for a range of encoded OAM values (top row) between [-3,3].}
\label{fig:ModeSortingillustration RESULTS}
\end{figure}

Figure \ref{fig:ModeSortingillustration RESULTS} experimentally illustrates the transformation performed by the mode sorter on LG modes ranging from -3 to 3 OAM. Here the top row shows the modes generated by the SLM and directed through the sorter. The OAM per photon is given above the respective spatial modes. In the row below, the elongated spots formed in the FP of the sorting lens are shown. Cross-hairs in the images mark the position of the 0 OAM mode. Consequently, the OAM dependent sorting power of the elements may be clearly seen where the negative OAM spots are formed to the left of the cross-hairs and the positive OAM spots to the right. Additionally, the position moves incrementally in the OAM handedness direction, based on the OAM value.

\subsubsection{Spot positions}

Quantitative evaluation of the spot positions was carried out whereby experimental measurement of the spots formed for varying OAM was compared to calculated values determined from Eq. \ref{eqn: spotPOS}. Anticipation of both the accuracy and consistency which may be expected for the modes detected with the elements is thus possible. The expected function of the distance between sorted modes relative to the OAM $(l)$ then
\begin{equation}
t_{r} = \frac{\lambda f l}{d} = \frac{(633 \text{ nm})(500 \text{ mm})}{(10.5 \text{mm})}l = (30.1 l \text{ }\mu\text{m}) \nonumber
\end{equation}
for the refractive sorter,
\begin{equation}
t_{r} = \frac{\lambda f l}{d} = \frac{(633 \text{ nm})(100 \text{ mm})}{(1.12 \text{mm})}l = (59.6 l \text{ }\mu\text{m}) \nonumber
\end{equation}
for 1-copy diffractive sorter and
\begin{equation}
t_{r} = \frac{\lambda f l}{d} = \frac{(633 \text{ nm})(100 \text{ mm})}{(0.5 \text{mm})}l = (126.6 l \text{ }\mu\text{m}) \nonumber
\end{equation}
for the 3-copy diffractive sorter.

\begin{figure}[h]
\centering
\includegraphics[width=\linewidth]{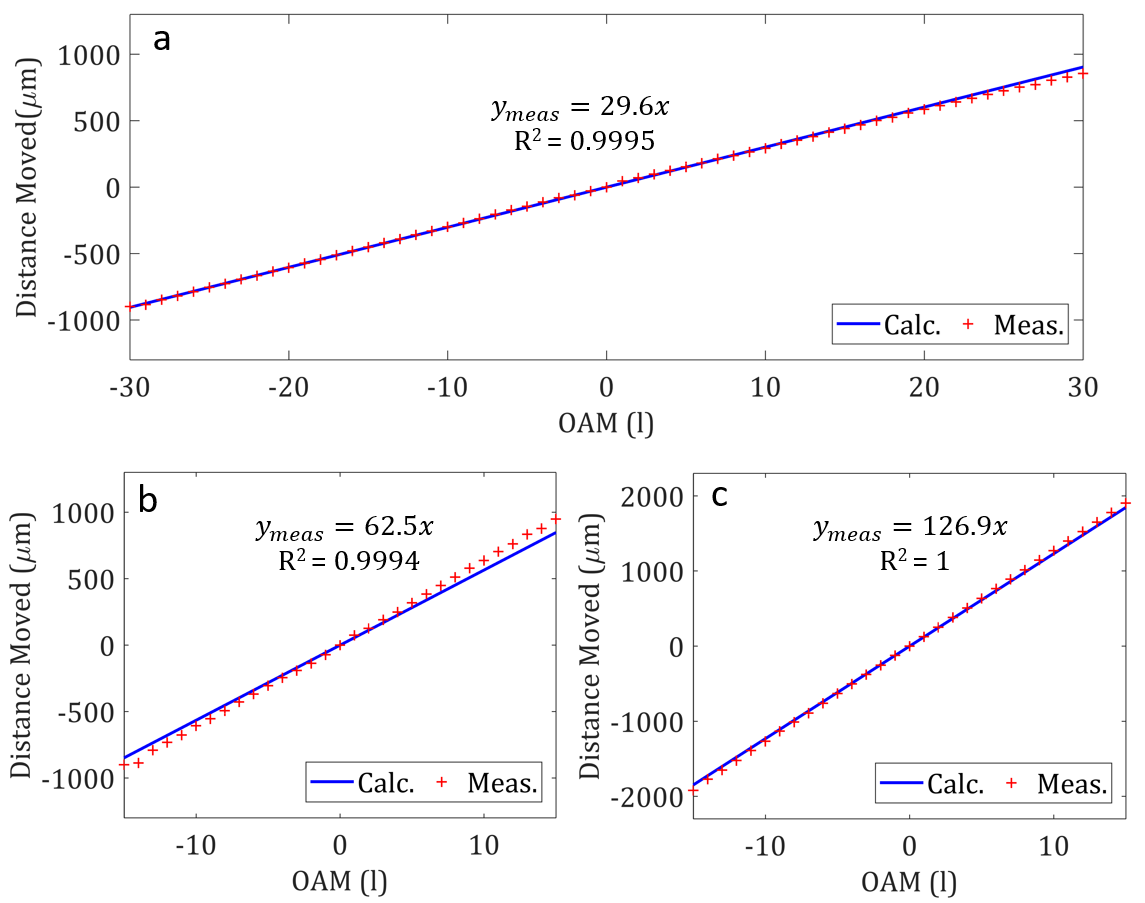}
\caption[Shift in sorted spot position with OAM]{A comparison of measured (red) sorted spot positions against that which was theoretically expected (blue) based on the designed unraveled beam length, d for the (a) refractive, (b) 1-copy and (c) 3-copy mode sorter.}
\label{fig:ModeSortingposition RESULTS}
\end{figure}

Results are given in Fig. \ref{fig:ModeSortingposition RESULTS} where the positions detected for the range of OAM values directed through the sorter are plotted alongside the calculated functions determined above. 

Here the positions were determined relative to the $l=0$ mode. Based on the previous discussion on the viability of OAM ranges, the measured OAM set was greater for the refractive sorter with OAM = [-30,30], than the diffractive sorter with OAM = [-15, 15].

Considering Fig. \ref{fig:ModeSortingposition RESULTS} (a), the experimental shift in spacing agrees well with the calculated value obtained from the design parameters with a difference of 30.1$\mu$m –- 29.6$\mu$m = 0.5$\mu$m. Additionally, the positions form a straight line as evidenced with the high $R^{2}$-value of 0.9995. This indicates a consistent positional shift which should lead to defined boundaries between the experimentally formed OAM spots. As a result, the detected intensities at those positions should be an accurate reflection of the modes and weightings in the beams being sorted.  

Based on the spacing calculations, the expected separation distance for the 1-copy mode sorter was 59.6 $\mu$m compared to the average experimental value of 62.5 $\mu$m in Fig. \ref{fig:ModeSortingposition RESULTS} (b). This deviation is notable with a 62.5 $\mu$m –- 59.6 $\mu$m = 2.9 $\mu$m difference as it accumulates with the number of increased modes as evidenced by in Fig \ref{fig:ModeSortingposition RESULTS} (b) for the higher OAM modes. A straight line, however, is still formed by the experimental shift in position with OAM. This may be explained by the length of the unwrapped beam where it may have been slightly shorter in the experimental implementation than the parameter quoted. The performance of the 1-copy sorter shall thus be evaluated with the experimental positional shift where adequate performance may be expected with a similarly high $R^{2}$-value of 0.9994 compared to the refractive sorter. Accordingly, the sorter performs adequately. 

For the 3-copy mode sorter, comparison of the calculated and measured values yielded a 126.9 $\mu$m –- 126.6 $\mu$m = 0.3 $\mu$m difference, indicating excellent agreement. Additionally, the $R^{2}$-value of 1 gives rise to good expectations in terms of detecting the correct OAM values.

\subsubsection{Alignment range and cross-talk}

\begin{figure}
\centering
\includegraphics[width=\linewidth]{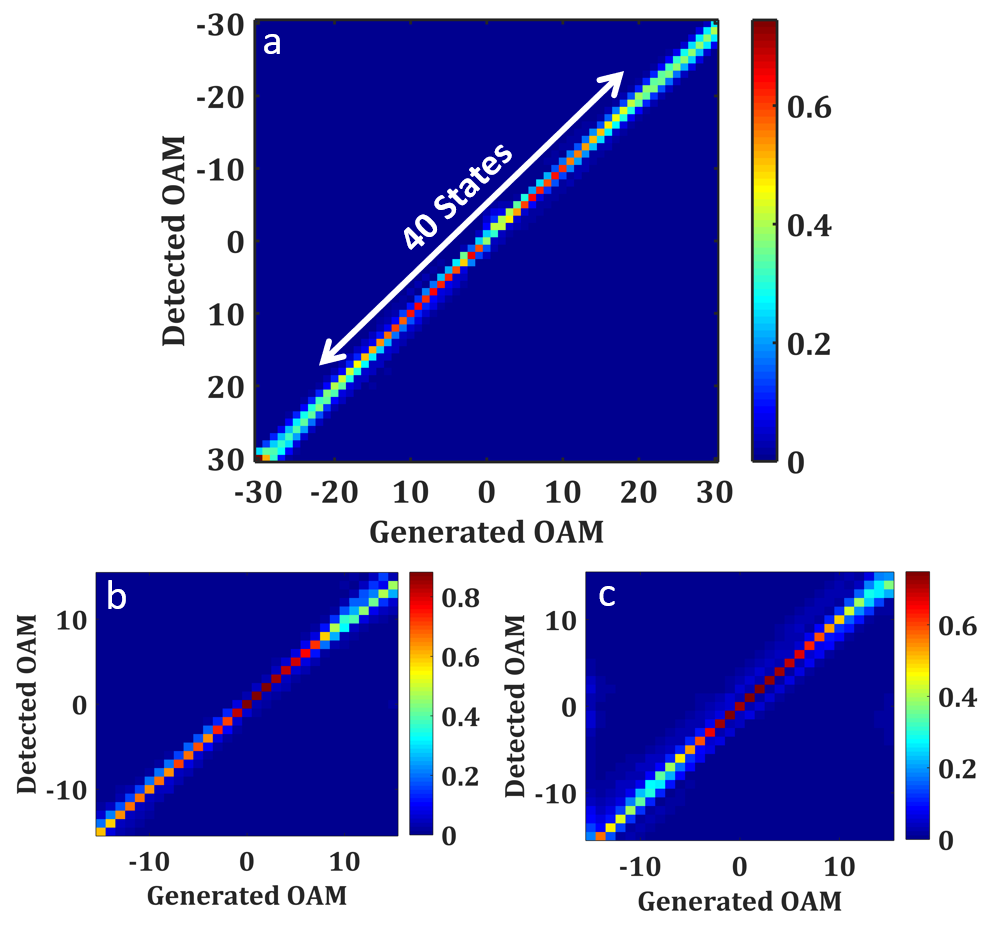}
\caption[Mode sorter OAM spectrum results]{A density plot showing the sorting power and the associated cross-talk that may be expected with acceptable alignment for the (a) refractive, (b) 1-copy mode sorter and (c) 3-copy mode sorter.}
\label{fig:ModeSortingCrosstalk RESULTS}
\end{figure}

By binning rectangular areas on the CCD along the direction of the spot movement, the intensity was integrated over to determine both the presence and weighting of different OAM modes. Here the bin positions were determined by Eq. \ref{eqn: spotPOS} with appropriate adjustment in the 1-copy case. Accordingly, determination of the viable range of OAM modes for which the MS may be used was achieved by binning for a range of OAM modes across the CCD image captured of each generated mode. OAM modes from -30 to 30 for the refractive sorter and -15 to 15 for the diffractive sorters were then sent separately through the MS and the detected modes then read out. The results are given in Fig. \ref{fig:ModeSortingCrosstalk RESULTS} for the different sorters.   

Here the presence of cross-talk is indicated by the off-diagonal elements. It may be seen that the minimum cross-talk achievable was ~25\% for the most aligned spot detected (as shown by the range of the color map) for both the refractive and 1-copy sorters in Fig \ref{fig:ModeSortingCrosstalk RESULTS} (a) and (b) respectively. It may also be observed that near 0 OAM, the refractive sorter generated a larger amount of cross-talk than in the diffractive mode sorter elements for the best alignments that were achievable. Figure \ref{fig:ModeSortingCrosstalk RESULTS} (c) yields the characteristic density plot for analysing the 3-copy mode sorter. Here the minimum cross-talk achievable may be seen as substantially reduced with only 10\% being detected in the incorrect modes.

Alteration of the strongest detected modes away from the diagonal indicated the OAM mode detected is no longer correct. As a result, the viable sorting range may be established by observing the number of modes where the maximum intensity remains within the diagonal. It follows from the discussion on the effect of the skew angle of the incident beam, that deviation in detected modes is expected to occur as the OAM-value increases. This should occur later in the refractive sorter case, however. The deviation is observed as a decrease in intensity of the diagonal values and a spreading in the off-diagonal terms as the generated OAM values digress from the central 0 OAM value.

In Fig. \ref{fig:ModeSortingCrosstalk RESULTS} (a), the range of modes for which correct OAM modes are detected is [-20,20] with the refractive sorter. From Fig. \ref{fig:ModeSortingCrosstalk RESULTS} (b), with the 1-copy sorter, about 15 modes fall within a range of acceptable accuracy where the intensity that is detected in the correct mode remains above 60\%. A greater range may still be used for the mode sorter, provided correction terms are used for the non-existent modes being detected as well as loss of weighting in the correctly detected modes. However, this is also limited as past 20 modes, the defining spot disintegrates into a fringe array for which the positions are not indicative of the modes present. 

The range of viable modes with detected weightings greater than 60\% for the correct mode increased to 23 in the 3-copy case. Additionally, the spread of the cross-talk between modes was reduced due to the additional number of unwrapped beams. A stronger along-side diagonal may be seen, however, in comparison to Figure \ref{fig:ModeSortingCrosstalk RESULTS} (b). This may be attributed to the greater misalignment between the elements as the sensitivity of the element increased with the number of the beams copied. Subsequently, additional fringes were caused directly adjacent to the spot, yielding a stronger presence of erroneous detection of adjacent OAM modes.

It follows that greater accuracy of the detected modes was found with the diffractive sorters, however, the range of OAM modes were diminished by the small radius of the element. As a result, the refractive sorter remained consistent over almost twice the OAM range in comparison to the diffractive elements, illustrating this point for a beam size of 6 mm. Increasing the incoming beam size will subsequently also increase the sorting range.

\subsubsection{Spot resolution}

\begin{figure}
\centering
\includegraphics[width=\linewidth]{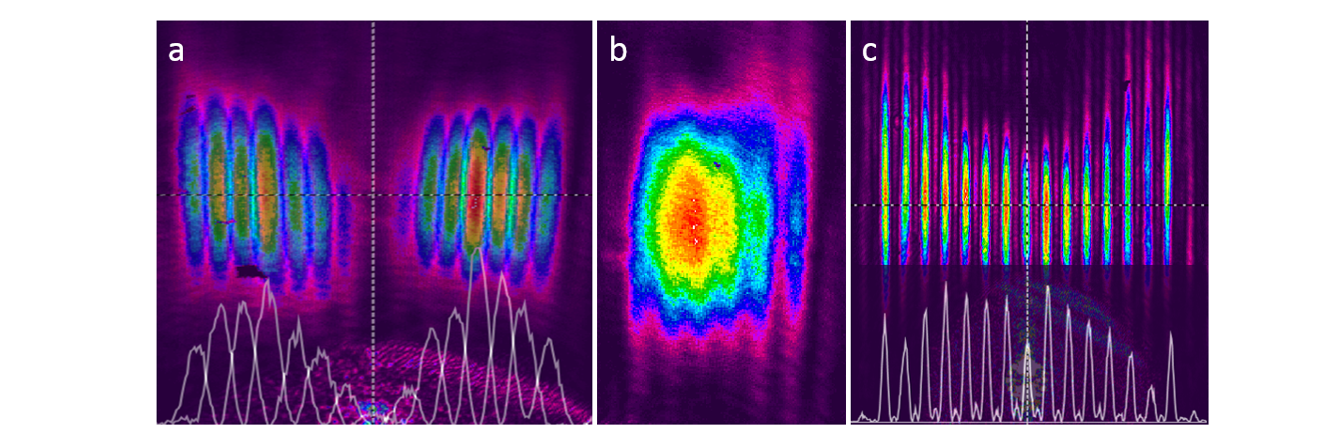}
\caption[Overlapping of sorted modes]{Images showing resolution for adjacent OAM modes when sorted with (a) a 1-copy mode sorter with the alternating odd and even modes between [-7,7] superimposed along with their 2D profiles, (b) adjacent OAM modes between [-1,-8] sorted by the 1-copy mode sorter and (c) adjacent OAM modes between [-7,7] sorted by the 3-copy mode sorter.}
\label{fig:ModeSortingOverlap RESULTS}
\end{figure}

By superimposing sets of alternating OAM spots, the resolution of adjacent spots were investigated. This was done for the 1-copy sorter in Fig. \ref{fig:ModeSortingOverlap RESULTS} (a).  Here superpositions of even and odd OAM modes in the interval [-7,7] were separately sent through the sorter and imaged. Superimposing these modes clearly describes how the modes overlap (also seen shown with the 2D profiles below the spots). A clear overlap may consequently be seen which will lead to additional cross-talk for the detection of OAM modes not present.

Additionally, a set of adjacent modes where OAM = [-1,-8] were sent through the mode sorter with the resulting distribution shown in Fig. \ref{fig:ModeSortingOverlap RESULTS} (b). The convolution resulted in a distortion of the intensity spectrum associated with the OAM present as well as eliminating the ability to visually distinguish a spot’s position. The latter may be evidenced through the appearance of only 6 spot ‘tails’ at the bottom of the convolution when 8 OAM modes are present.

The 3-copy mode sorter, however, effectively separated and defined the spots for adjected OAM modes as is illustrated for Fig. \ref{fig:ModeSortingOverlap RESULTS} (c). Here a superposition of adjacent OAM modes [-7,7] was generated the corresponding detected spots shown in the figure. A 2D profile is shown below the spots, clearly illustrating the reduction in overlap and increased spot resolution. 

\subsubsection{Weighted detection}

\begin{figure}[b]
\centering
\includegraphics[width=\linewidth]{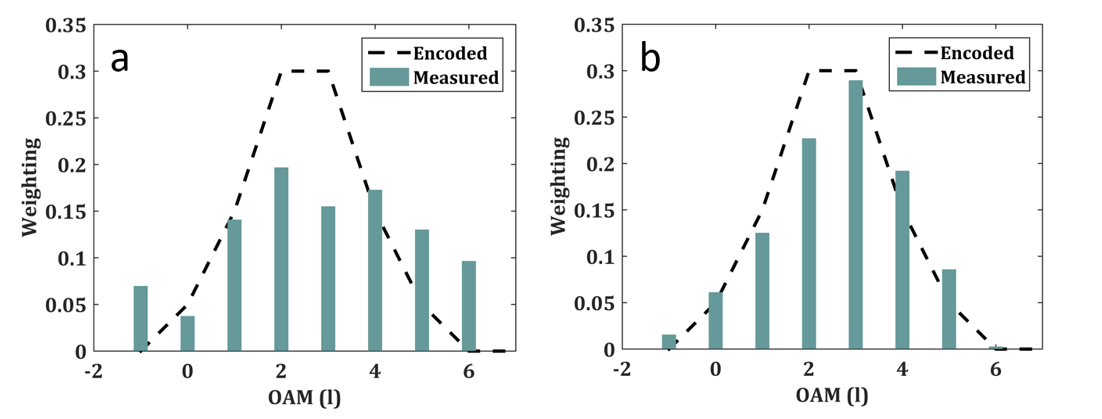}
\caption[Mode weighting detection assessment]{Relative weightings evaluated for multiplexed OAM modes sent through the mode sorter for the (a) 1-copy and (b) 3-copy mode sorters with respective similarities of S = 0.791 and S = 0.968.}
\label{fig:ModeSortingWeighting RESULTS}
\end{figure}

Accurate detection of mode weightings was also evaluated with the results given in Fig. \ref{fig:ModeSortingWeighting RESULTS} (a) and (b) for 1- and 3-copy sorters respectively. Here a distinct superposition of OAM modes were multiplexed by the SLM and sent through the sorters.

Comparative accuracy of the multiplexed and detected OAM modes was determined through its similarity as given by, 
\begin{equation}
S = \frac{[\sum_{l}\sqrt{W_{exp}(l)W_{th}(l)}]^{2}}
{\sum_{l}W_{exp}(l)\sum_{l}W_{th}(l)}
\label{eqn: Simularity}
\end{equation}
where $W_{th}(l)$ is the theoretical or multiplexed weighting associated with the OAM mode l and $W_{exp}(l)$ is the detected equivalent of the mode. Observation of the Fig. \ref{fig:ModeSortingWeighting RESULTS} (a) and (b) indicate that the multiplexed and detected weightings resemble each other, showing that either of the sorters would be suitable for detection, however, a significant increase in the accuracy of modal detection occurs as the copy numbers increase. More specifically, the similarity of 79.1\% for the 1-copy is increased by 22\% when adding 2-copies.

Consequently, both the refractive and diffractive sorters exhibited advantages and disadvantages associated with their implementation. Specifically, the refractive mode sorter was more robust, allowing a greater range of beam sizes to be easily sorted as well as maintaining a larger range of OAM values when generating larger input beam sizes. The accuracy associated with measuring the weightings associated with a 1-copy sorter, such as the refractive one, however, is adversely affected with a 79.1\% accuracy which may be expected if reasonable alignment is achieved. Here, the 3-copy diffractive sorter is more advantageous with a large increase in accuracy with fair alignment. In addition, the cross-talk measured was substantially smaller with a reduction in the power erroneously detected for incorrect OAM modes. Implementation of this sorter resulted in a higher sensitivity to misalignment, resulting in a more difficult detection system as well as a significant restriction of the size of the beam that can be sorted. Furthermore, variation of beam sizes that may be sorted is small as only a deviation of 0.300mm is possible in comparison to the 8mm range achievable in the refractive sorter case. Subsequently, for more robust requirements, the refractive sorter was favorable at the expense of the system accuracy; conversely, greater accuracy was achieved with the 3-copy diffractive sorter at the expense of the range and some stability of the detection system.

\subsection{Experimental considerations and data analysis}

\subsubsection{Pulse overlap and adjusting the prediction}

The time gap $(t)$ between each output pulse from the resonator is related to the length of the resonator $(L)$ through the relation, $t=L/c$ where $c$ refers to the speed of light in air and t the time. Here, the resonator must be longer or equal to the temporal pulse width to avoid overlap of the circulating pulse and thus an overlap in the QW steps. However, experimental consideration as well as stability and alignment factors resulted in an upper limit restriction of 3 m for the resonator perimeter. It was subsequently designed in a $0.3 m \times 1.7 m$ rectangular configuration.

Measurement of the subsequent temporal pulse width of the laser was carried out by placing a Thorlabs DET210-a photodiode before the ICCD and averaging over 10 intensity pulses on a Tektronix TDS2024B 1GHz oscilloscope. The corresponding temporal parameters were then estimated by fitting the pulse to a Gaussian function of the form:
\begin{equation}
G(x) = a e^{-(x-b)^{2}/(2c^{2})}+k,
\label{eqn:Gaussmodel}
\end{equation}
 where a determines the height of the pulse function, x refers to the $x$-axis position which is then adjusted by $b$. $c$ indicates the width of the pulse and $k$ adjusts the position along the $y$-axis. The corresponding parameters extracted from fitting Eq. \ref{eqn:Gaussmodel} are summarized in Table \ref{table:fitparam}
A SSE (sum of the squares of the error) of 0.002573 and R-squared value of 0.974 determined from the fitting both indicate that the function as well as the parameters are adequate reflections of the measured pulse for the walker.

\begin{table}[h]
\centering
\caption[Pulse fitting]{\bf{Parameters for Gaussian fit to average pulse length measured from the Spectrophysics laser.}}
\label{Table:ModeSortingParameters}
\begin{tabular}{m{0.3\columnwidth}>{\centering\arraybackslash}m{0.3\columnwidth}}
\toprule
\textbf{Parameter} & \textbf{Fitted Value}\\
\hline
\textit{a} & 0.06045 \\
\hline
\textit{b} & 0.106 x $10^{-9}$ \\
\hline
\textit{c} & 6.107 x $10^{-9}$\\
\hline
\textit{k} & 0.0040\\
\hline
\end{tabular}
\label{table:fitparam}
\end{table}

\noindent Here the parameters may subsequently be used to yield both the mathematical description of the pulse as well as indicate the full-width half maximum (FWHM) and the full-width at one-tenth maximum (FWTM) values for the pulse length. 

The FWHM is then given by
\begin{equation}
FWHM = 2\sqrt{2\ln{(2)}} \times c = 14.3ns \nonumber
\end{equation}

with the FWTM also determined by
\begin{equation}
FWTM = 2\sqrt{2\ln{(10)}} \times c = 26.6ns \nonumber
\end{equation}

As the parameter $b$, indicates the position of the peak relative to a position space, $b = 0$ allows the function to be centred at the origin, making it easier to work with. Additionally, $k$ indicates the $y$-position which is also irrelevant, hence $k = 0$. It follows that the intensity distribution of the pulse may be characterized by the Gaussian function, $G(t$):
\begin{equation}
G(t) = 0.0605 e^{-t^{2}/(2(6.107 ns)^{2})}
\label{eqn:Gaussfit}
\end{equation}

It follows that the FWHM is close to the 10 ns resonator length with 14.3 ns. Comprehensive prediction of the mode spectrum may be determined by considering the FWTM value, however, which is almost 3 times larger than the 10 ns expected for the resonator design, indicating a significant overlap. Correction for this was achieved by modelling of the OAM mode spectrum overlap.

\begin{figure}
\centering
\includegraphics[width=\linewidth]{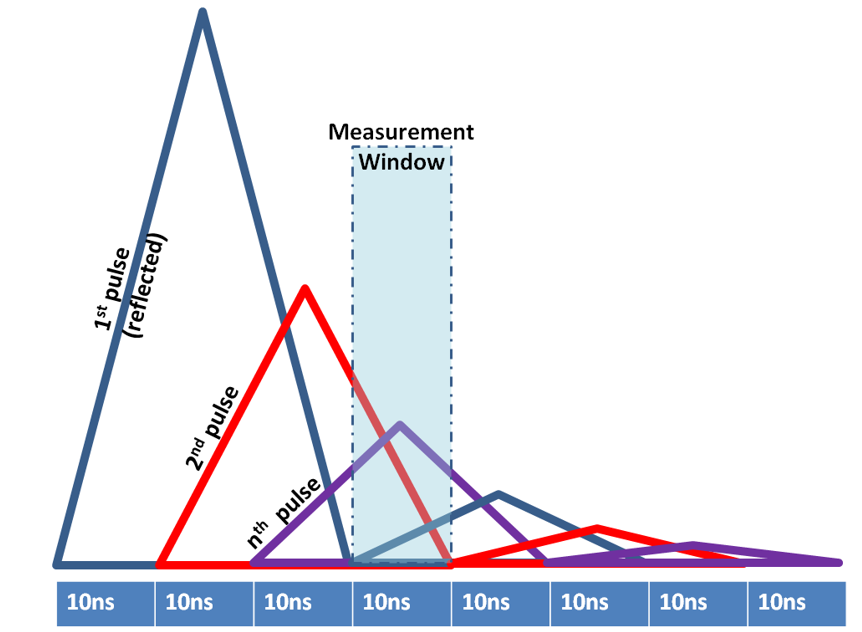}
\caption[Measuring spatial modes in a QW]{Illustration of pulse overlap and intensity effects for each pulse emitted from the resonator or step in the quantum walk.}
\label{fig:measurementwindow}
\end{figure}

Two factors were considered in the experimental simulation of the step distribution expected for each round trip. These factors were (1) how the pulses overlap and (2) the intensity decrease occurring each RT due to the partial transmission through the BS.  More specifically, a FWTM of ~30 ns and resonator length of 10 ns,the output pulse will contain contributions from the previous pulse as well as the next pulse. This is illustrated in Fig. \ref{fig:measurementwindow}.

As the measurements were taken every 10 ns for each respective step, the overlap was minimized as indicated by the blue dotted box. The second factor regarded the decrease in the internal resonator intensity which lead to a difference in the intensities of the overlapping components. Again, this is illustrated in Fig \ref{fig:measurementwindow}. Subsequently, the correction factor for the ‘previous’ pulse overlap contribution was larger than that of the ‘future pulse’. A general formula for a transmission percentage $T$ and subsequent reflection, $R = (1-T)$, of the pulse may easily be derived. This is given in Eq. \ref{eqn:weighing}

\begin{equation}
w(n) = \Bigg \{ \begin{matrix}
            0  & if \: n < 0 \\
 			R & if \: n = 0 \\
 			T^{2}R^{n-1} & if \: n \neq 0
\end{matrix}
\label{eqn:weighing}
\end{equation}

The temporal shape of the pulse additionally affects the overlap between the pulses. This was approximated by the best fit curve for the average temporal pulse as determined in Eq. \ref{eqn:Gaussfit}. These intensity and temporal width factors were used to augment the simulated QW.

\begin{figure} [b]
\centering
\includegraphics[width=\linewidth]{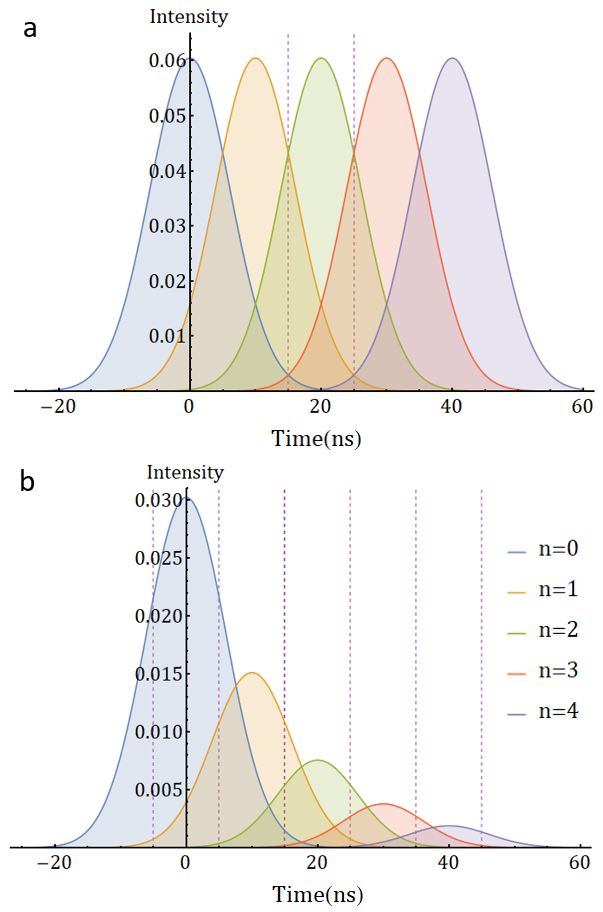}
\caption[Overlap as pulses lose intensity]{Adjusted diagram illustrating the pulse overlap for a 3 m = 10 ns resonator configuration where (a) the overlap is emphasized and (b) the weighting effects associated with loss in intensity incorporated.}
\label{fig:PulseOverlapIntRed}
\end{figure}

An adjusted diagram of the pulse overlap is presented as Fig. \ref{fig:PulseOverlapIntRed} (a) and (b) below. Figure \ref{fig:PulseOverlapIntRed} (a) indicates the extent of the pulse overlap while (b) illustrates the effects of the different weightings for a $50:50$ beamsplitter.

\begin{figure}[h]
	\centering
	\includegraphics[width=\linewidth]{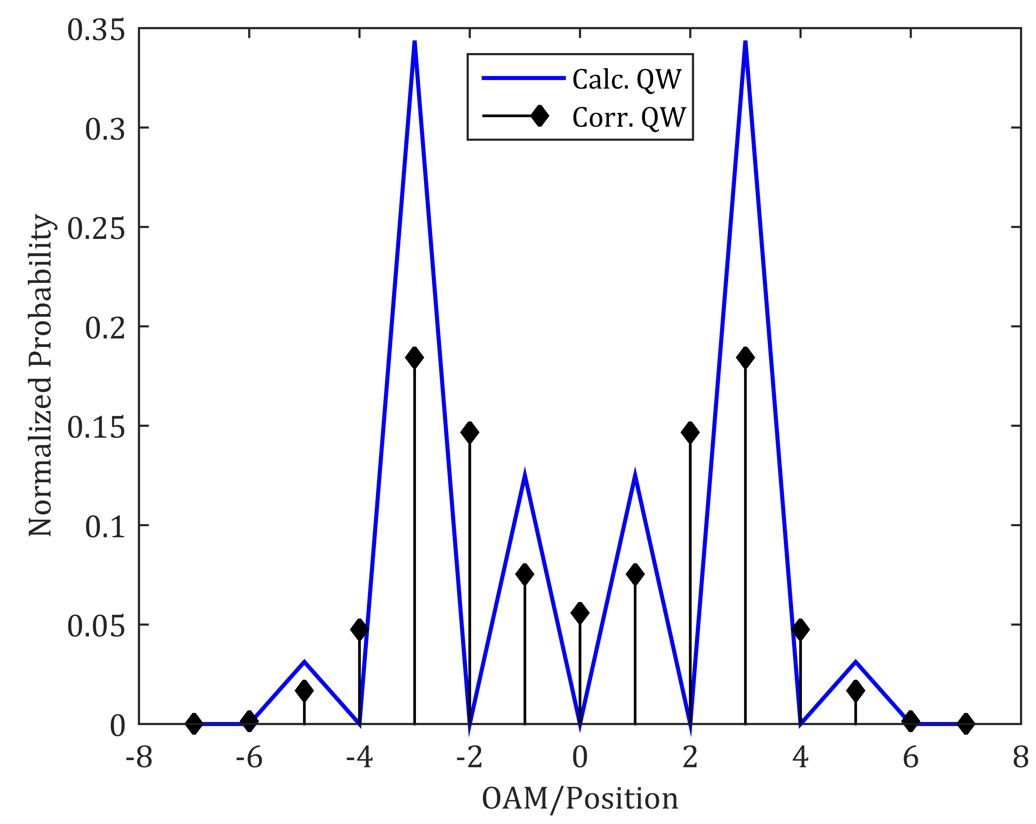}
	\caption[Corrected QW distribution]{Step 5 of the OAM QW distribution with (Corr. QW) and without (Calc. QW) the experimental corrections for step, $n = 5$ or detected pulse 6.}
	\label{fig:QWcorrecstep5}
\end{figure}

The dotted lines indicate the section of the pulse that was gated by the AndOR camera for the $nth$ step in the quantum walk series. This gated section of the pulse was determined using the formula:
\begin{equation}
a = \frac{PW-GW}{2}
\label{eqn:capturedsection}
\end{equation}
where PW is the complete pulse width (40 ns based on the modelled pulse) and GW is the gate width (i.e. the time section of pulse captured).Based on the AndOR gating time, the pulse was captured from 
\begin{align}
&\frac{-PW}{2} + a = -20 +(\frac{40-10}{2}) = -5 \nonumber\\
&\text{  to  } \nonumber \\
&\frac{PW}{2} - a = 20 -(\frac{40-10}{2}) = 5. \nonumber
\label{eqn:outer limits capture}
\end{align}
Correction to the quantum walk probability distribution for the 1st pulse $(n=0)$ was thus:
\begin{align}
&\text{\textbf{n = 0:}} \nonumber \\
&[[QW_P(0)] w(0)\int^{\frac{PW}{2} - a}_{\frac{-PW}{2} + a} G(t)dt] \nonumber \\
&+ [[QW_P(1)] w(1)\int^{\frac{-PW}{2} + a}_{\frac{-PW}{2} + a -10} G(t)dt] \nonumber \\
&+[[QW_P(2)] w(2)\int^{\frac{-PW}{2} + a - 10}_{\frac{-PW}{2} + a - 20} G(t)dt]  \nonumber
\end{align}
Similarly, for $n$ = 1 and 2:
\begin{align}
&\text{\textbf{n = 1:}} \nonumber \\
&[[QW_P(0)] w(0)\int^{\frac{PW}{2} - a+10}_{\frac{PW}{2} - a} G(t)dt] \nonumber \\
&+[[QW_P(1)] w(1)\int^{\frac{PW}{2} - a}_{\frac{-PW}{2} + a} G(t)dt] \nonumber \\
&+[[QW_P(2)] w(2)\int^{\frac{-PW}{2} + a}_{\frac{-PW}{2} + a - 10} G(t)dt] \nonumber \\
&+ [[QW_P(3)] w(3)\int^{\frac{-PW}{2} + a-10}_{\frac{-PW}{2} + a-20} G(t)dt] \nonumber
\end{align}
\begin{align}
&\text{\textbf{n = 2:}} \nonumber \\
&[[QW_P(0)] w(0)\int^{\frac{PW}{2} - a+20}_{\frac{PW}{2} - a+10} G(t)dt] \nonumber \\
&+ [[QW_P(1)] w(1)\int^{\frac{PW}{2} - a+10}_{\frac{PW}{2} - a} G(t)dt] \nonumber \\
&+[[QW_P(2)] w(2)\int^{\frac{PW}{2} - a}_{\frac{-PW}{2} + a} G(t)dt] \nonumber \\
&+ [[QW_P(3)] w(3)\int^{\frac{-PW}{2} + a}_{\frac{-PW}{2} + a-10} G(t)dt] \nonumber \\
&+[[QW_P(4)] w(4)\int^{\frac{-PW}{2} + a-10}_{\frac{-PW}{2} + a-20} G(t)dt]. \nonumber
\end{align}
Accordingly, the following general model applies:
\begin{equation}
\begin{split}
	&\text{\textbf{n:}} \\
	&[[QW_P(n-2)] w(n-2)\int^{\frac{PW}{2} - a+20}_{\frac{PW}{2} - a+10} G(t)dt]\\
	&+[[QW_P(n-1)] w(n-1)\int^{\frac{PW}{2} - a+10}_{\frac{PW}{2} - a} G(t)dt]\\
	&+[[QW_P(n)] w(n)\int^{\frac{PW}{2} - a}_{\frac{-PW}{2} + a} G(t)dt]  \\
	&+ [[QW_P(n+1)] w(n+1)\int^{\frac{-PW}{2} + a}_{\frac{-PW}{2} + a-10} G(t)dt] \\
	&+[[QW_P(n+2)] w(n+2)\int^{\frac{-PW}{2} + a-10}_{\frac{-PW}{2} + a-20} G(t)dt], 
	\end{split}
	\label{eqn:pulseoverlap}
\end{equation}
where $QW_P(n)$ is the quantum walk probability distribution of the $nth$ step. For comparison, Fig. \ref{fig:QWcorrecstep5} shows the normalized expected and corrected Hadamard probability distributions for the $6th$ pulse $(n=5)$. 
The corrected distribution displayed is for the case of a $50:50$ beam-splitter acting as the resonator output window.
Comparing the calculated distribution to that of the corrected one, the same diverging trend is occurs. The maxima here still appear near the edges of the distribution with the highest weightings occurring at the same positions. It follows that the expected difference occurring from the overlapping modes within the cavity is a broadening of the peaks as well as the presence of probabilities in the adjacent positions. Consequently, the generated QW still retain the characteristic qualities associated with the pure QWs.

\subsubsection{Beam profile and spatial filtering}

\begin{figure}
\centering
\includegraphics[width=\linewidth]{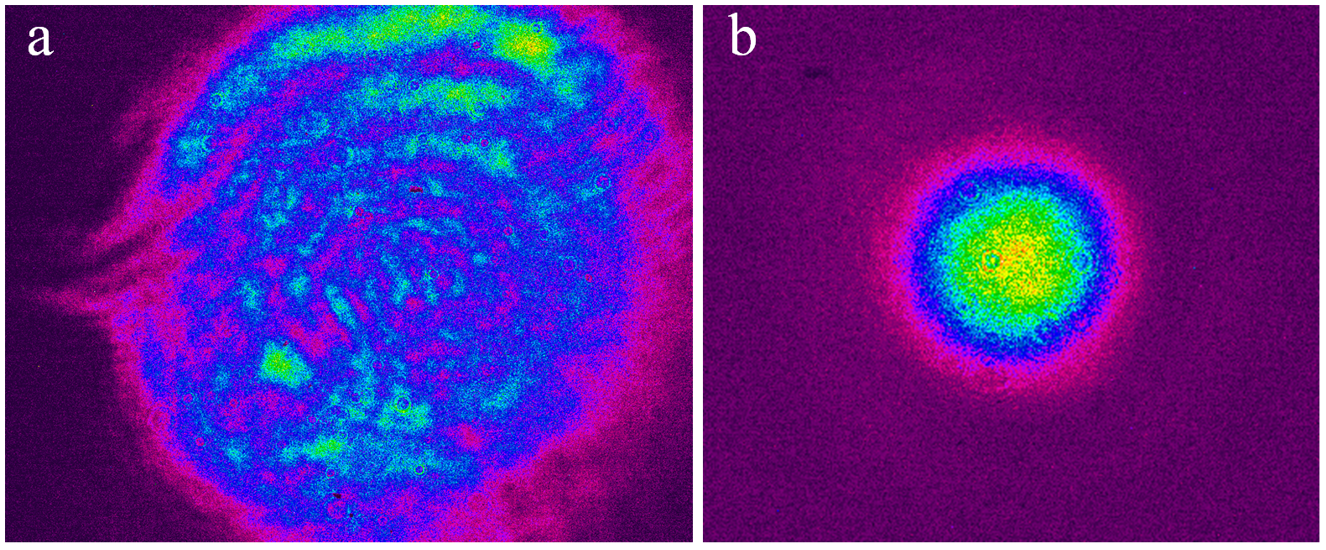}
\caption[Correcting laser beam profile]{False color map images of the near-field pulsed laser output beam (a) before spatial filtering and (b) after spatial filtering. The spatially filtered beam is smaller than the original as it was de-magnified in the spatial filtering process.}
\label{fig:CorrProfile}
\end{figure}

Elimination aberrations and unwanted modes contained in the initially laser generated beam was necessary through construction of a spatial filter. Figure \ref{fig:CorrProfile} (a) shows the transverse output profile generated by the Spectraphysics DCR-11 laser. From the transverse distribution, the presence of aberrations and additional modes are evident in addition to a large beam width of 3 mm. A spatial filter was implemented with a 50 $\mu m$ pinhole placed in the Fourier plane of the $4-f$ system with lenses of focal lengths, $f1 = 750$ mm and $f2 = 200$ mm respectively. Here the pinhole was three times larger than the calculated Gaussian beam width in the FP, resulting in the extraneous modes being filtered out spatially. The focal length ratio ($f2 / f1$) between the lenses additionally formed a demagnification telescope to reduce the beam diammeter to 2 mm.

The spatial filter output beam is given in Fig. \ref{fig:CorrProfile} (b) where a Gaussian intensity distribution can be seen along with the appropriate demagnification. In addition, the reduced size allowed for a more sustainable walk with the dimensions remaining below the size of the optics.

\subsubsection{Pulse synchronization and gating}

\begin{figure}[h]
\centering
\includegraphics[width=\linewidth]{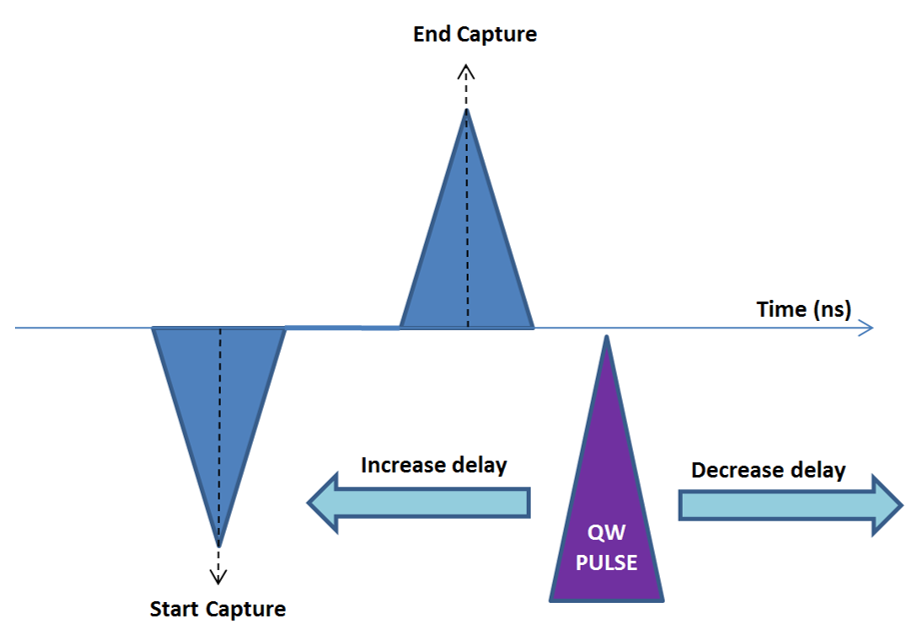}
\caption[Synchronization]{Schematic illustrating the procedure necessary to correctly capture an output pulse for a single step or round trip in the quantum walk.}
\label{fig:Sync}
\end{figure}

An AndOR iStar ICCD camera (734 series) with a photocathode optical shutter allowed gating times at the 10 ns scale and thus isolation of the correct round tip. The opening and closing of the photocathode was determined by monitoring the electronic signals, sent to enact the operation in the camera, with an oscilloscope. As indicated in Fig. \ref{fig:Sync}, the opening is determined with a negative pulse and closing with a positive pulse.

\begin{figure}
\centering
\includegraphics[width=\linewidth]{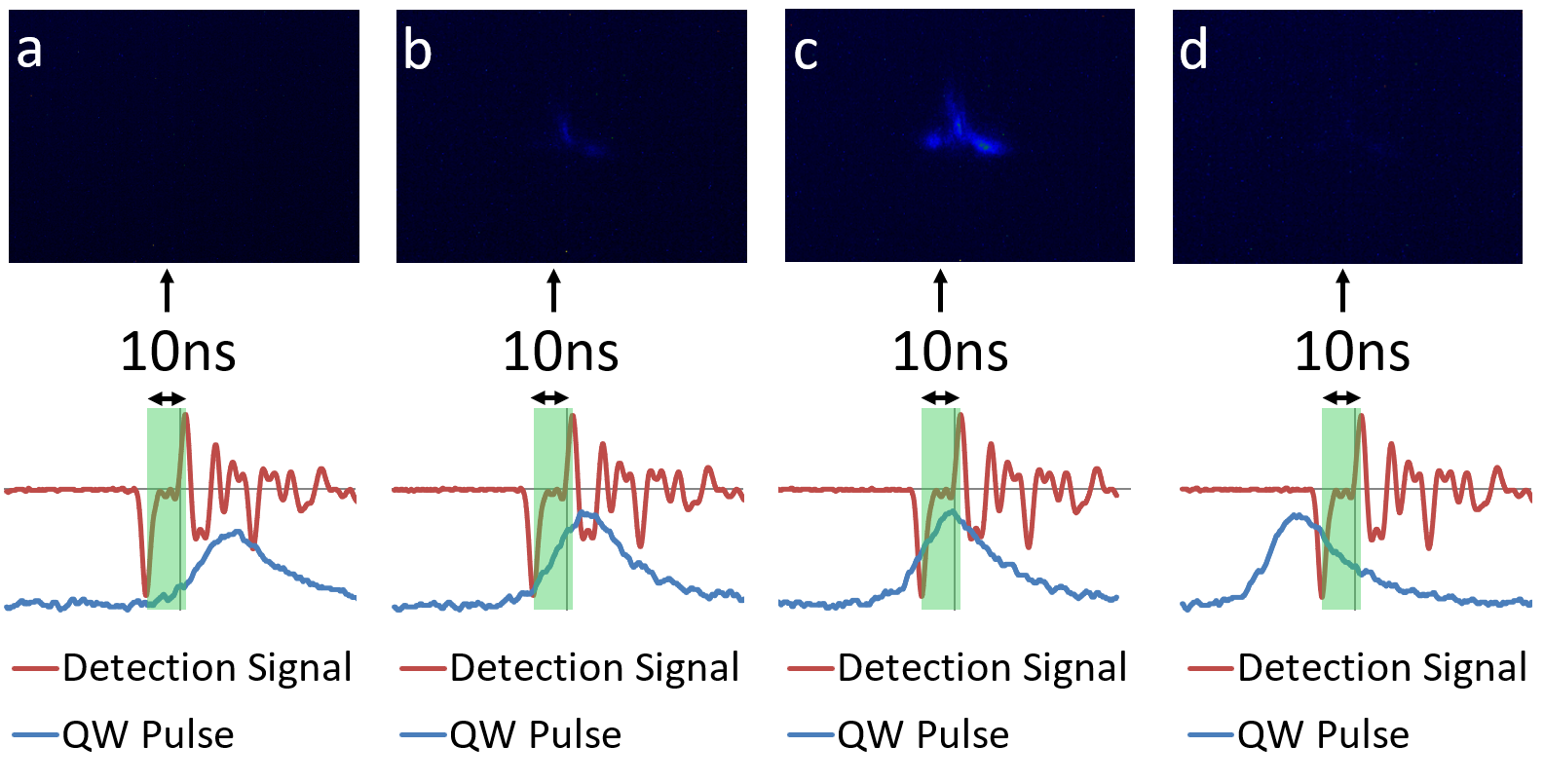}
\caption[Synchronization with triggering]{Experimental illustration of synchronization required between the AndOR camera and pulsed laser for accurate output pulse capture from the QW resonator. Delay time settings were decreased from (a-d).}
\label{fig:SyncTrig}
\end{figure}

The camera and laser pulse emission was synchronised with a digital delay generator so as to locate the desired output pulse (step). A specific time delay was then placed between the pulse emission and opening of the photocathode. The delay chosen was equivalent to the time taken for the electronic signals to travel to the respective components as well as the time for the pulse to reach the resonator, circulate the resonator a desired number of times, travel through the sorter and reach the camera. To determine the delay, a photodiode was placed at the same distance from the output as the camera. Arrival of the pulse once fired triggered the oscilloscope which was also monitoring the photocathode trigger signal in a second channel. Adjustment of the calculated time delay altered the first detected pulse (step 0) temporal position as depicted in Fig. \ref{fig:SyncTrig}. After determination of the initial pulse delay, isolation of the desired step or pulse output was achieved by adding an additional delay of $Nt$ that is related to the resonator circulation time, $t$ and number of steps, $N$.

Determination of the appropriate delay required for synchronization of the camera and laser pulse is demonstrated in Fig. \ref{fig:SyncTrig}. The graphs along the lower row of the figure are signals recorded by a oscilloscope monitoring the photocathode triggering signals as well as the arrival of the emitted laser pulse due to a trigger pulse from the digital delay generator for the same nanosecond time scale ($x$-axis). The $y$-axis then represents the associated voltages. It follows that the blue profile is the laser pulse detected by the photodiode, placed the same distance as the camera from the output pulse and the red profile is the trigger signals operating the opening (first large dip) and closing (first large spike) of the photocathode. Transparent green overlays on the respective graphs emphasize the period for which the camera is recording the pulse intensity and thus the section of the pulse captured. As indicated, the gate width (capturing time) was set to 10 ns.

Images above each of the oscilloscope graphs are the transverse intensity distributions captured in the respective 10 ns gated windows where the delay between the photocathode and laser triggers from the digital delay generator was systematically decreased from (a) through to (b) for a three-lobe spatial profile. It can thus be seen from (a)-(b) how the pulse moves into the capturing window of the camera before the maximum pulse intensity is captured in (c) and then moving past the range again in (d). The variation in intensity of the captured modes clearly show the synchronization effects of the time delay parameter and the desired position for ideal analysis of an output pulse with (c).

\begin{figure}
\centering
\includegraphics[width=\linewidth]{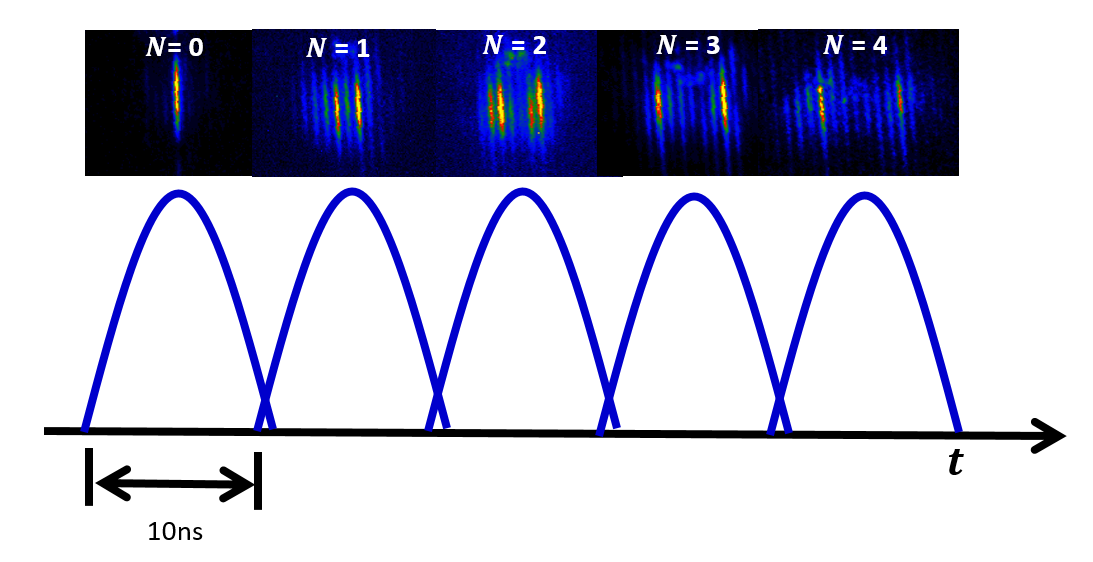}
\caption[Illustration of change in OAM with time]{Example evolution of output intensity distributions as captured every consecutive 10ns from the first output pulse, marking each round trip made by the circulating light beam for the QW resonator.}
\label{fig:QW illustration}
\end{figure}

Application of the appropriate synchronization time delay as determined with the method illustrated in Fig. \ref{fig:SyncTrig} for the first pulse in the resonator setup is allowed for the capture of the desired QW step or output pulse. Subsequent addition of a step-related time delay constant, $Nt$, to the initial delay allowed for the capture of the intensity distribution related to that step ($N$) or output pulse. This is illustrated in Fig. \ref{fig:QW illustration} for the experimental setup. Here successive output pulse intensity distributions from $N = [0,4]$ were captured with a 10 ns gate width in the mode sorter Fourier plane. A clear evolution in the distribution may be seen across the steps, indicating a spread in the intensity distribution and thus the successful capture of additional output pulses according to the associated delay settings. It follows that this method would be effective in attaining the OAM spectra of each QW step, allowing real-time observation of the walk.

\subsection{Experimental data correction}

\begin{figure*}
\centering
\includegraphics[width=\linewidth]{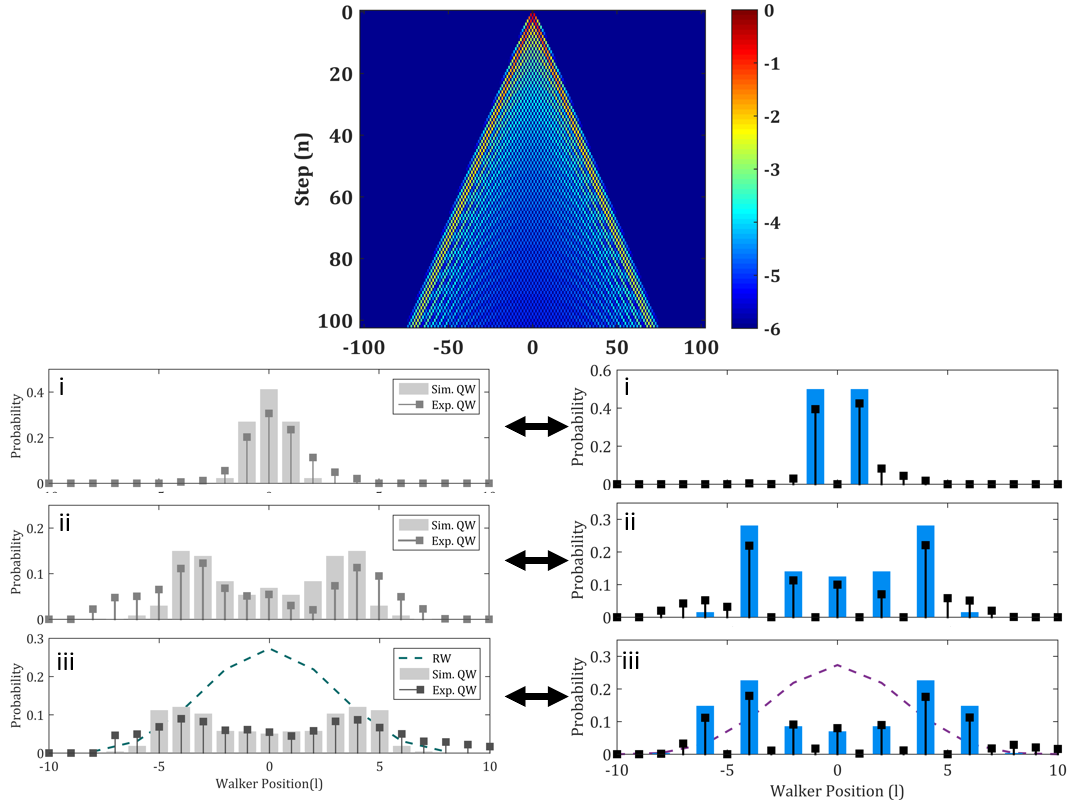}
\caption[Illustration of directly measured and experimentally corrected symmetrical Hadamard QW]{Illustration of the results presented in the paper for the symmetrical Hadamard QW with a simulated density plot indicating the traditional spread in distribution with a logarithmic color scale over 100 steps. Inset graphs below indicate correlation of the results where the theory was corrected for overlap (left) vs. where the experimental results were corrected for the overlapping pulses (right). Black lines (color bars) indicate the experimental (simulated) results for (i) Step 1, (ii) Step 6 and (iii) Step 8. The dotted line indicates the analogous Random walk distribution expected in the classical case for the last step.}
\label{fig:SymHadCORR}
\end{figure*}

\begin{figure*}
\centering
\includegraphics[width=\linewidth]{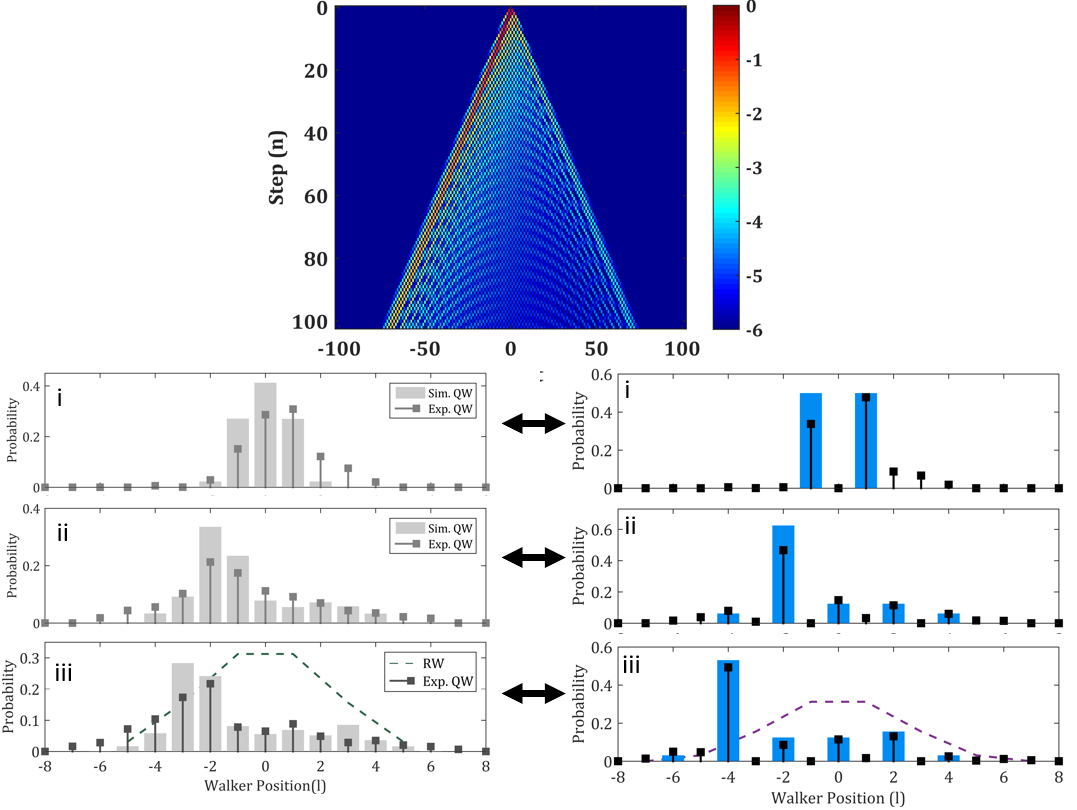}
\caption[Illustration of directly measured and experimentally corrected asymmetrical Hadamard QW]{Illustration of the results presented in the paper for the asymmetrical Hadamard QW with a simulated density plot indicating the traditional spread in distribution with a logarithmic color scale over 100 steps. Inset graphs below indicate correlation of the results where the theory was corrected for overlap (left) vs. where the experimental results were corrected for the overlapping pulses (right). Black lines (color bars) indicate the experimental (simulated) results for (i) Step 1, (ii) Step 4 and (iii) Step 5. The dotted line indicates the analogous Random walk distribution expected in the classical case for the last step.}
\label{fig:AsymHadCORR}
\end{figure*}

\begin{figure*}
\centering
\includegraphics[width=\linewidth]{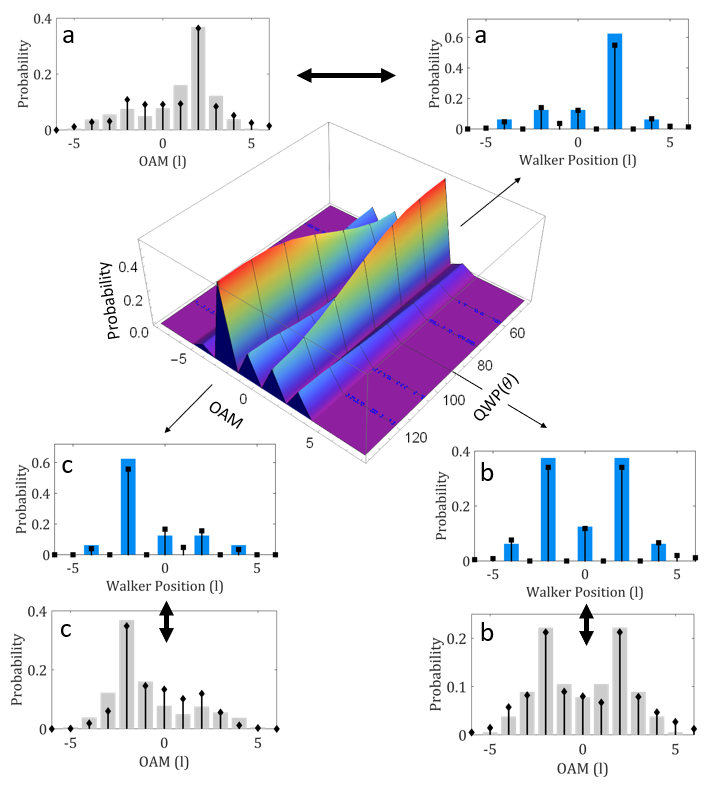}
\caption[Rotating the coin comparison]{Comparison of the change in QW symmetry presented in the paper for measured (black) and predicted distributions (bars) with and without correction for pulse overlap. Here the 3D plot indicates the traditional predicted distribution (without overlapping pulse correction) and the inset graphs show the alterations between correcting the theoretical distribution (gray) vs. correcting the experimentally measured distribution (blue) for QWP coin fast axis orientations of (a) $45^{\circ}$, (b) $90^{\circ}$ and (c) $135^{\circ}$}
\label{fig:QWPCORR}
\end{figure*}

\begin{figure*}
\centering
\includegraphics[width=\linewidth]{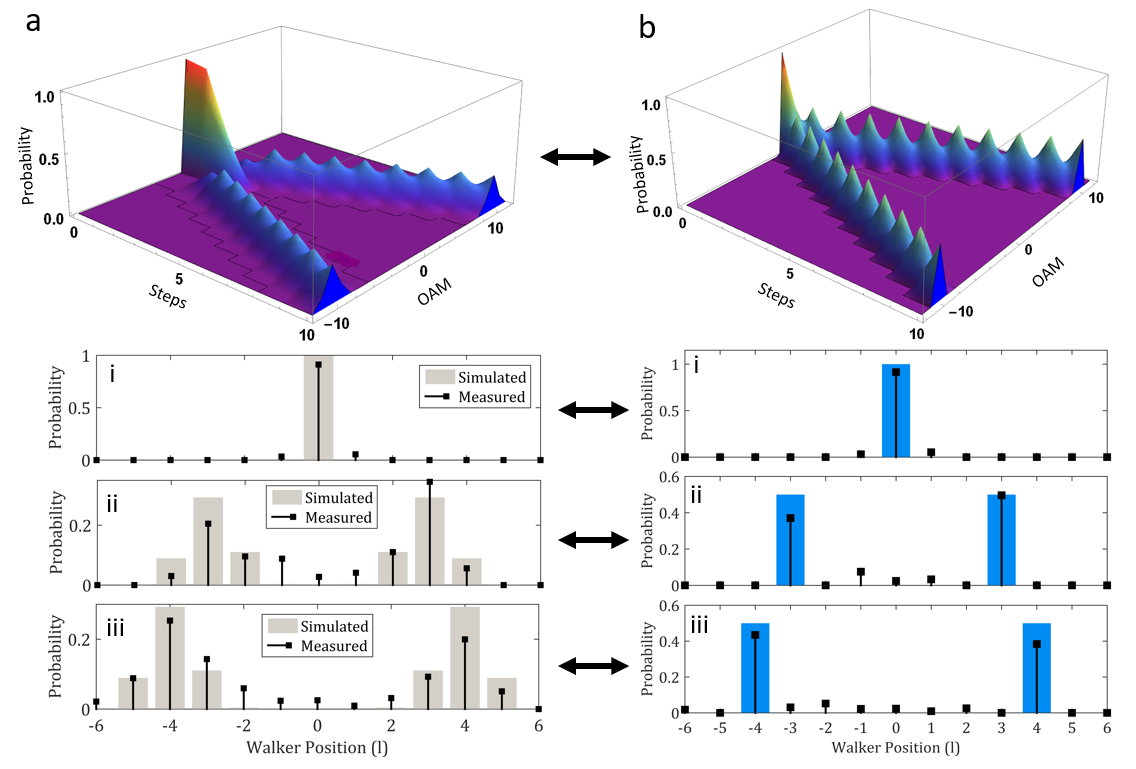}
\caption[Rotating the coin comparison]{Illustration of the results presented in the paper for the Identity coin QW with a simulated 3D plot indicating the traditional spread in distribution (left) and adjusted spread (right) due to pulse overlap. Inset graphs below indicate correlation of the results where the theory was corrected for overlap (left) vs. where the experimental results were corrected for the overlapping pulses (right). Black lines (color bars) indicate the experimental (simulated) results for (i) Step 0, (ii) Step 3 and (iii) Step 4.}
\label{fig:IDCORR}
\end{figure*}

\begin{figure*}
\centering
\includegraphics[width=\linewidth]{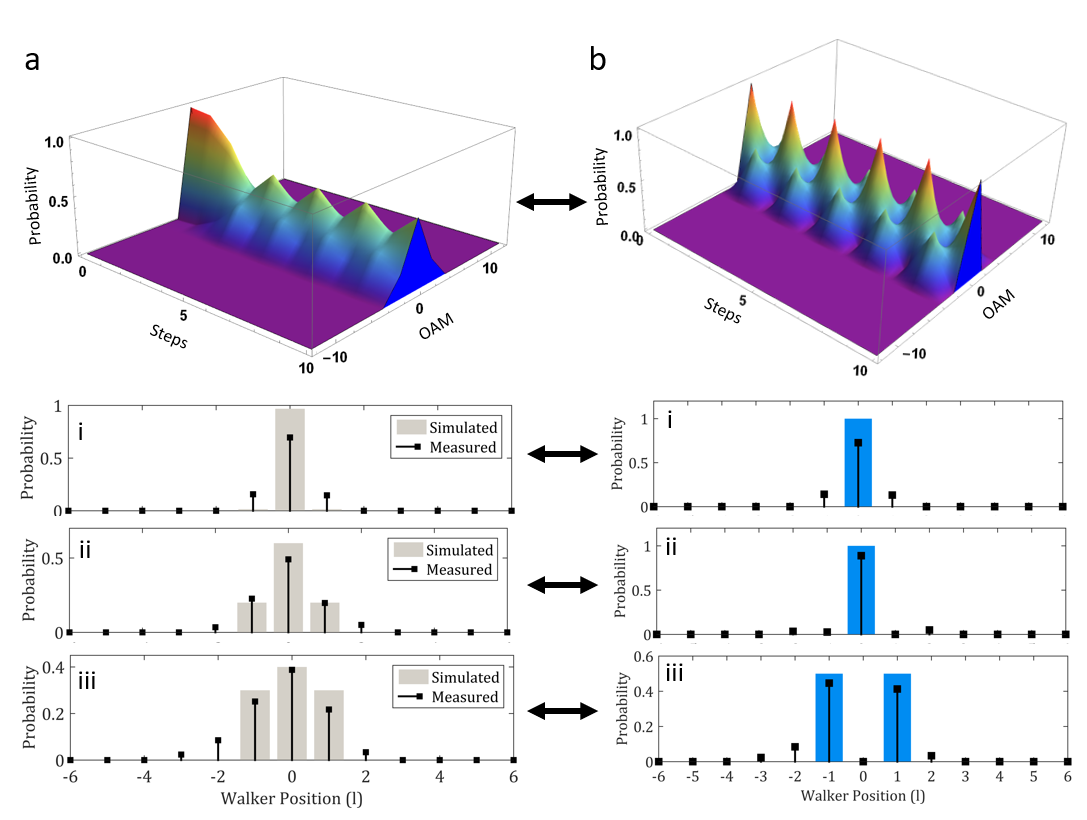}
\caption[Rotating the coin comparison]{Illustration of the results presented in the paper for the NOT-coin QW with a simulated 3D plot indicating the traditional spread in distribution (left) and adjusted spread (right) due to pulse overlap. Inset graphs below indicate correlation of the results where the theory was corrected for overlap (left) vs. where the experimental results were corrected for the overlapping pulses (right). Black lines (color bars) indicate the experimental (simulated) results for (i) Step 0, (ii) Step 4 and (iii) Step 5.}
\label{fig:NOTCORR}
\end{figure*}

Following the noted overlap occurring in the experimental setup and the associated derivation of the correction to the theory, the measured experimental results were expected to have an adjusted distribution as indicted by Eq. \ref{eqn:pulseoverlap}. Accordingly, the simulated distribution was altered as illustrated by Fig. \ref{fig:QWcorrecstep5}. Comparison between the directly measured experimental distribution and altered theory thus allowed the QW distribution to evaluated. The results in this form are shown by the gray bar graphs (to the left) in Fig. \ref{fig:SymHadCORR} for the symmetric Hadamard QW case, Fig. \ref{fig:AsymHadCORR} for the asymmetric Hadamard QW, Fig. \ref{fig:QWPCORR} where the QW symmetry was changed with the QWP fast axis orientation, Fig. \ref{fig:IDCORR} for the Identity coin QW and Fig. \ref{fig:NOTCORR} for the NOT-coin QW. 

However, in order appropriately evaluate the QW distributions with respect to what is traditionally expected, it was also possible to reverse the overlap and correct the measured distribution such that it reflected the characteristic QW commonly seen. This was accomplished by taking the QW distribution measured for the $nth$ step and applying the reverse of Eq. \ref{eqn:pulseoverlap} such that the $QW_{P}(n)_{meas}$ could be retrieved or de-convoluted from the measured distribution. 

To generate the experimental correction, consider simplifying the equation by simplifying the terms specified in Table \ref{table:coeffs} by the variables listed.
\begin{table}[h]
\centering
\caption[Equation term simplification]{\bf{Equation \ref{eqn:pulseoverlap} term simplification.}}
\label{Table:ModeSortingParameters}
\begin{tabular}{m{0.15\columnwidth}>{\centering\arraybackslash}m{0.5\columnwidth}}
\toprule
\textbf{Variable} & \textbf{Equation term}\\
\hline
$c_{(n-2)}$ & $w(n-2)\int^{\frac{PW}{2} - a+20}_{\frac{PW}{2} - a+10} G(t)dt]$ \nonumber\\
\hline
$c_{(n-1)}$ & $w(n-1)\int^{\frac{PW}{2} - a+10}_{\frac{PW}{2} - a} G(t)dt]$ \nonumber\\
\hline
$c_{n}$ & $ w(n)\int^{\frac{PW}{2} - a}_{\frac{-PW}{2} + a} G(t)dt]$ \nonumber\\
\hline
$c_{(n+1)}$ & $w(n+1)\int^{\frac{-PW}{2} + a}_{\frac{-PW}{2} + a-10} G(t)dt]$ \nonumber\\
\hline
$c_{(n+2)}$ & $w(n+2)\int^{\frac{-PW}{2} + a-10}_{\frac{-PW}{2} + a-20} G(t)dt]$ \nonumber\\
\hline
\end{tabular}
\label{table:coeffs}
\end{table}

Equation \ref{eqn:pulseoverlap} then becomes

\begin{align}
QW_{PCorr}(n) = & c_{(n-2)}QW_P(n-2) + c_{(n-1)}QW_P(n-1)\nonumber \\
& + c_{n}QW_P(n) + c_{(n+1)}QW_P(n+1) \nonumber \\
& + c_{(n+2)}QW_P(n+2),
\label{eqn:pulseoverlapv2}
\end{align}
\noindent where $QW_{PCorr}(n)$ is the probability distribution measured for the $nth$ step as a result of the overlapping pulses in the resonator.

Now in order to correlate the measured distribution to the expected, they should both be normalized by dividing the distribution by the sum i.e. $S = \sum{QW_{PCorr}(n)}$ and $S_{Exp} = \sum{QW_{P}(n)_{meas}}$. Letting $QW_{P}(n)_{NORMmeas} = QW_{P}(n)_{meas}/S_{Exp}$, it follows that 

\begin{align}
QW_{P}(n)_{NORMmeas} = & \Big[c_{(n-2)}QW_P(n-2) + c_{(n-1)}QW_P(n-1)\nonumber \\
& + c_{n}QW_P(n) + c_{(n+1)}QW_P(n+1) \nonumber \\
& + c_{(n+2)}QW_P(n+2)]\Big] \div S.
\label{eqn:pulseoverlapv2}
\end{align}

As we are now modifying the experimentally measured data, it follows that $QW_P(n)$ is becomes the corrected experimental data. Subequently,

\begin{align}
QW_P(n)_{meas} = &\frac{QW_{P}(n)_{NORMmeas} \times S}{c_{n}} - \nonumber \\ &\Big[c_{(n-2)}QW_P(n-2) + c_{(n-1)}QW_P(n-1)\nonumber \\
& + c_{(n+1)}QW_P(n+1) + c_{(n+2)}QW_P(n+2)]\Big] \div c_{n}.
\label{eqn:pulseoverlapv2}
\end{align}

After applying the correction, any miscellaneous negative values were taken as background errors and equated to 0 as a negative probability is not physically possible. The resulting distributions were then normalized and are respectively presented in the inset graphs to the right in Fig. \ref{fig:SymHadCORR} to Fig. \ref{fig:NOTCORR}. Here the blue bars indicate the traditional distributions expected for these types of QWs and no overlap adjustment is shown. This may be clearly seen by where the uncorrected measured values occupy adjacent OAM states or positions while the corrected results shown occupy alternate OAM states or positions. The double sided arrows in each case indicate the interchangeable corrections between the simulated and measured probability distributions and the subsequent matching correlations between the experimental and simulated distributions.

\end{document}